\documentclass{emulateapj}
\usepackage{graphicx}
\usepackage{amsmath}

\usepackage{color}\def\red#1{\textcolor{black}{#1}}

\newcommand{\msun}{\, M_\odot}
\newcommand{\mlim}{m_{\rm limit}}
\newcommand{\mth}{m_{\rm th}}
\newcommand{\Mbh}{M_{\rm BH}}
\newcommand{\Mpc}{\rm Mpc}
\newcommand{\hMpc}{h^{-1}{\rm Mpc}}

\shorttitle{}
\shortauthors{Komiya et al.}

\begin{document}
\title{A Cross-correlation Analysis of AGN and Galaxies using Virtual Observatory: Dependence on Virial Mass of Super-Massive Black Hole} 

\author{Yutaka Komiya, 
Yuji Shirasaki, 
Masatoshi Ohishi, and 
Yoshihiko Mizumoto
}

\begin{abstract}
We present results of the cross-correlation analysis between active galactic nuclei (AGNs) and galaxies at redshift 0.1--1. 
We obtain data of $\sim 10,000$ SDSS AGNs in which their virial masses with a super-massive black hole (SMBH) were estimated. 
The UKIDSS galaxy samples around the AGNs were obtained using the virtual observatory. 
The scale length of AGN-galaxy cross-correlation for all of the samples is measured to be $r_0= 5.8^{+0.8}_{-0.6}\hMpc$ (for the fixed slope parameter $\gamma=1.8$). 
We also derived a dependence of $r_0$ on the BH mass, $\Mbh$, and found an indication of an increasing trend of $r_0$ at $\Mbh>10^8\msun$. 
It is suggested that the growth of SMBH is mainly driven by interactions with the surrounding environment for $\Mbh>10^8\msun$. 
On the other hand, at $\Mbh\lesssim10^8\msun$, we did not find the BH mass dependence. 
This would imply that, for less massive BHs, the mass growth process can be different from that for massive BHs. 
\end{abstract}


\section{Introduction}\label{introS}
It is thought that most galaxies harbor supermassive black holes (SMBHs) in their centers, and that gas accretion onto a SMBH is the energy source of active galactic nucleus (AGN) \citep[e.g.][]{Richstone98, Ferrarese05}. 
There are strong correlations between mass of a SMBH and observational properties of its host galaxy such as velocity dispersion and stellar mass of the bulge \citep[e.g.][]{Magorrian98, Ferrarese00}. 

Growth of the BH mass is strongly coupled to the evolution of the host galaxy, but the process of ``co-evolution" is still unknown. 
It is thought that BHs grow by accretion of gas and/or merger of BHs, but the physical mechanisms of gas inflow toward central regions of galaxies and coalescence of binary BHs are not yet revealed. 
In the standard hierarchical structure formation framework, major mergers of galaxies should have played important roles for evolution of galaxies, growth of SMBHs, and AGN activity. 
Thus, investigating the clustering of the galaxies around AGNs is crucial to understand evolution of SMBHs and galaxies. 

Recent large-scale surveys, such as the Sloan Digital Sky Survey (SDSS), provide observational sample over $100,000$ AGNs \citep{Schneider10}. 
The auto-correlation function of AGNs was studied using the SDSS sample \citep{Shen07, Shen09, Ross09}. 
The clustering of galaxies around AGNs in the areas of deep surveys was also investigated by some authors \citep[e.g.][]{Coil09, Mountrichas09}. 
They showed that the cross-correlation function between AGNs and galaxies is similar to the auto-correlation of luminous red quiescent galaxies. 
\citet{Hickox09} found that radio selected AGNs are strongly clustered, and that infra-red selected AGNs are weakly clustered than optically selected AGNs.  
Recently, cross-correlation between AGNs and galaxies was studied using large samples. 
\citet{Donoso10} computed cross-correlation between $\sim$14,000 radio-loud AGNs at $z=0.4$--0.8 and reference luminous galaxy sample, and compared the clustering amplitude of radio-galaxies with that of $\sim 7,000$ SDSS quasars. 
They argued that radio-loud AGNs are clustered more strongly than radio-quiet ones. 
\citet{Krumpe12} investigated clustering of galaxies around $\sim3,000$ X-ray selected AGNs and $\sim8,000$ optically selected AGNs at $z<0.5$. 
No significant difference was found between X-ray selected and optically selected broad-line AGNs. 

In order to understand interaction between growth of SMBHs and their surrounding environment, it is important to investigate dependence of clustering amplitude on physical properties of BHs. 
\citet{Shen09} studied the dependence of the two-point auto-correlation function of quasars on luminosity, BH mass, color, and radio loudness, and found weak or no dependence on virial BH mass. 
The cross-correlation function between AGNs and galaxies will achieve smaller uncertainties in clustering measurement because it has many more pairs at a given separation compared with the auto-correlation function of AGNs. 
\citet{Donoso10} found positive dependence of the cross-correlation amplitude on stellar mass $M_*$, but their sample is radio-loud AGNs within a narrow range of stellar mass ($10^{11}\msun < M_* < 10^{12}\msun$). 
There is no previous study to investigate dependence of the cross-correlation function between AGNs and galaxies on BH mass over a wide mass range. 

In this study, we investigate dependence of clustering amplitude of galaxies around AGNs on the BH mass to reveal a relation between mass growth of SMBHs and large scale environment of their host galaxies. 
We have collected observational data of a large number of AGNs and galaxies by using the Japanese Virtual Observatory (JVO)\footnote{http://jvo.nao.ac.jp/portal/}, and computed the cross-correlation function between AGNs and galaxies. 
Thanks to the large sample covering a large area of the sky, we were able to perform the clustering analysis in a manner free from the cosmic variance. 
To investigate BH mass dependence over a wide mass range ($\gtrsim2 {\rm dex}$), we also use the sample of less massive BHs by \citet{Greene07} in addition to the SDSS quasar catalog \citep{Shen11}, as described below. 

\red{ Throughout this paper, we use the following cosmological parameters: $\Omega_M=0.3$, $\Omega_\Lambda=0.7$, $h=0.7$. 
All magnitudes are given in the AB system.
All of the distances are measured in comoving coordinates.
}

\section{Datasets}\label{dataS}
The AGN sample was extracted from two catalogs by \citet{Shen11} and \citet{Greene07}, both of which contain the estimated virial mass, $\Mbh$, of the SMBHs. 
\citet{Shen11} derived virial mass of BHs for 105,783 quasars in the SDSS DR7 quasar catalog~\citep{Schneider10}. 
About a half of their samples were uniformly selected by the criteria described in \citet{Richards02} and the remaining samples were selected by a variety of earlier algorithms or serendipitous selections. 

\citet{Greene07} derived BH masses for $\sim 8,500$ active galaxies at $z<0.35$ based on SDSS DR4 spectra. 
They also analyzed spectra for objects classified as galaxies, not only quasars. 
They extracted the AGN components from spectra of galaxies, and estimated BH masses. 
Their catalog contains more objects than the sample of \citet{Shen11} for $\Mbh \lesssim 10^7 \msun$. 

\begin{figure}
\includegraphics[width=\linewidth]{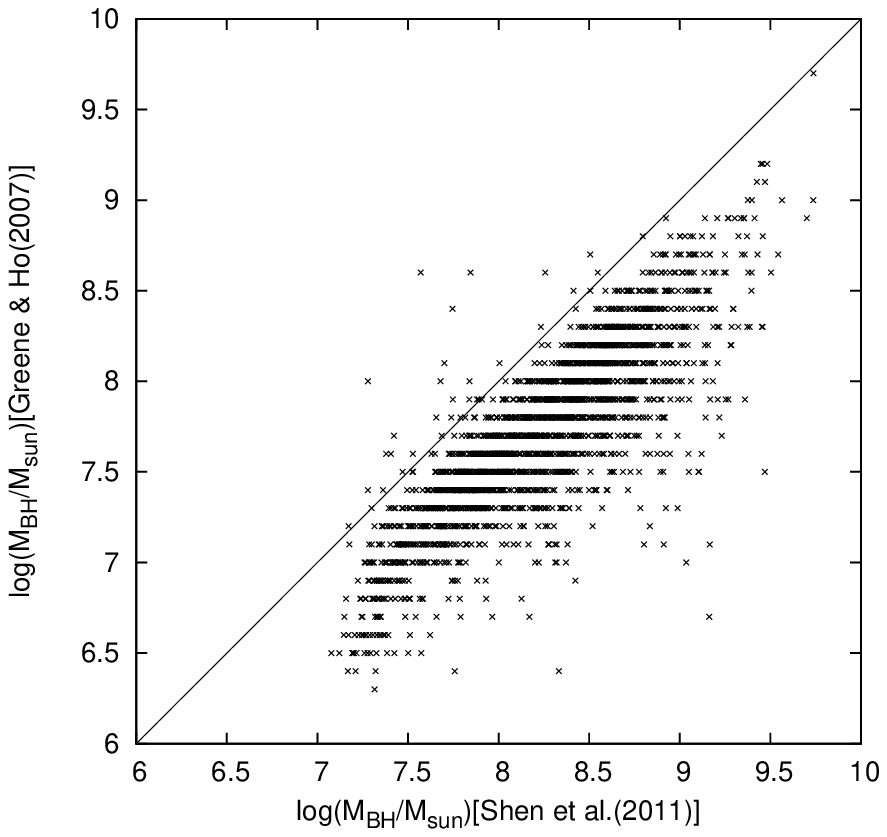}
\includegraphics[width=\linewidth]{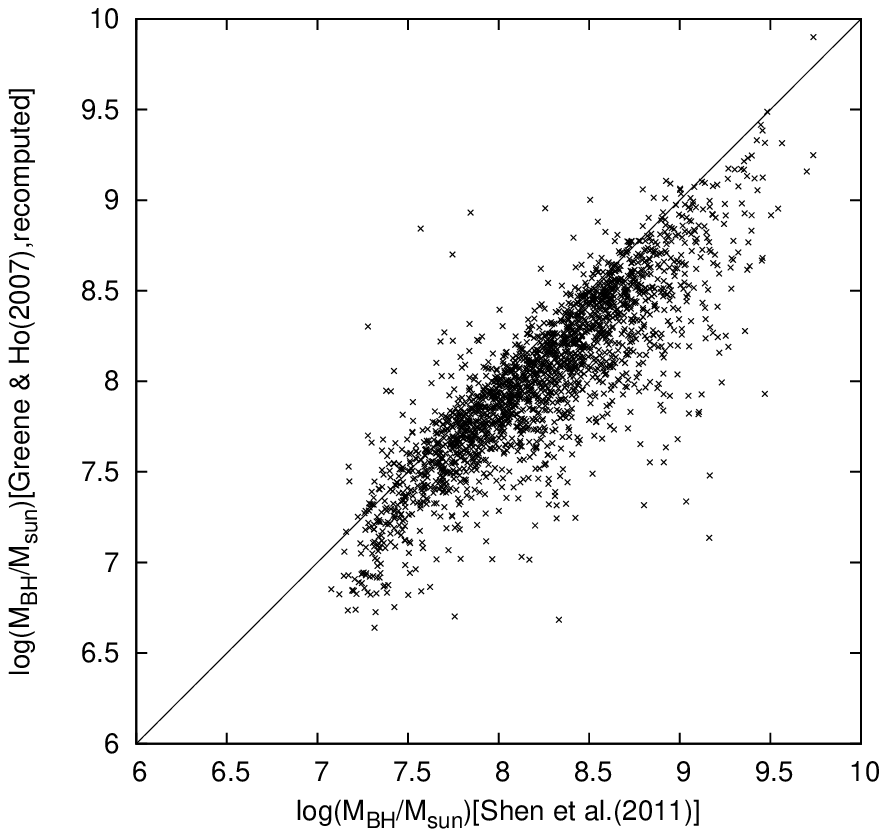}
\caption{
(Top panel)Black hole masses which is registered in the catalogs of \citet{Shen11} and of \citet{Greene07}. 
2,139 objects are registered in the both catalogs. 
There is a mean offset of $\sim0.5$ dex. 
(Bottom panel)For the data of \citet{Greene07}, black hole masses were recomputed using the FWHM and luminosity in the catalog of \citet{Greene07} and the parameter values in the virial mass estimator in \citet{Shen11}. 
Offset and scatter decrease but a mean offset of $\sim0.2$ dex remains. 
See text for detail. 
}\label{mass-mass}
\end{figure}

They both estimated virial mass of BHs by means of observed FWHM of the emission lines of H$\alpha$, H$\beta$ or Mg~II and the continuum luminosity at the lines. 
\citet{Shen11} used H$\beta$ estimates for $z<0.7$ and Mg~II estimates for $z\geq0.7$ as a fiducial virial mass estimate but also gave estimates based on other lines in their catalog when the lines were detected.  
For the samples of \citet{Shen11}, we used their fiducial mass estimate in this paper. 
\citet{Greene07} used H$\alpha$ estimates. 
We have found that there is systematic difference of $\sim0.5$ dex between the virial masses estimated in the two catalogs mainly because they used different parameter values in the virial mass estimator. 
Figure~\ref{mass-mass} shows estimated mass of 2,139 AGNs which are registered in the both catalogs. 
For the samples of \citet{Greene07}, we have recomputed virial mass by means of the FWHM and luminosity in \citet{Greene07} with the parameter values in \citet{Shen11} for H$\alpha$ line. 
For the recomputed virial mass, the systematic difference is decreased as shown in the bottom panel of Figure~\ref{mass-mass}. 
The recomputed masses by means of the data of \citet{Greene07}, however, are still $\sim 0.2$ dex smaller than the estimated masses in the catalog of \citet{Shen11} on average. 
A half of this remaining systematic difference is because of that FWHM and luminosity values were derived from the H$\alpha$ line for \citet{Greene07} but from H$\beta$ line for \citet{Shen11}. 
The other half is because of that \citet{Greene07} measured luminosity and FWHM of the emission lines of the extracted AGN components but \citet{Shen11} measured the lines without eliminating the host galaxy component from the spectra. 
However, this difference with $\sim0.2$ dex do not change our main conclusions as shown later. 
\red{For the AGNs registered in the both two catalogs, we use data in the catalog of \citet{Shen11}. }

We used the UKIDSS DR8 Large Area Survey (LAS) catalog \citep{Lawrence12} for galaxy samples. 
For each AGN, UKIDSS $K$-band data are searched around the AGN coordinates within 1 degree. 
We selected AGN samples of $z>0.1$ in order to analyse projected number density as a function of 
projected distance $r_p$ within $r_p \le 7$Mpc in the following. 
To obtain the data around the sample AGNs, we repeat to search galaxy data in the UKIDSS LAS catalog to the number of the sample AGNs. 
We obtained the data by accessing to the UKIDSS VO service through the JVO command line tools. 
By means of the VO tools we can recurrently access large data archives easily.  
To remove stars from the UKIDSS samples, we selected data for which the merged class flag 
equals to 1 (galaxies) or $-$3 (probable galaxies).
We also removed data of poor quality which have the post-processing error quality 
bit flags larger than 255.

\begin{figure}
\plotone{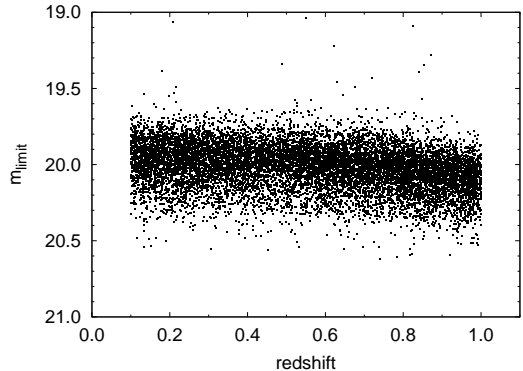}
\caption{Distribution of the \red{$K$-band} limiting magnitude $\mlim$ of the UKIDSS galaxies sample
as a function of AGN redshift. 
Each point denote $\mlim$ for the area around each AGN sample.  
11,335 AGN samples of $z=0.1$ -- 1.0 are distributed in the survey area of UKIDSS LAS. }
\label{fig:mlimit-z}
\end{figure}

\begin{figure}
\plotone{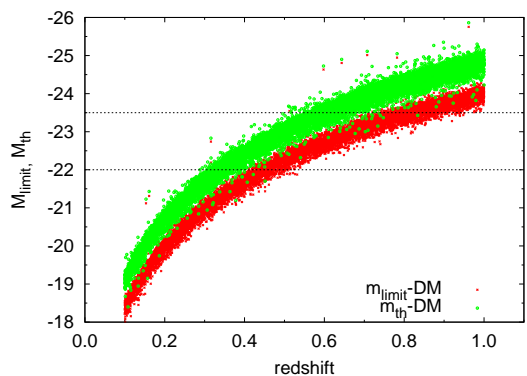}
\caption{Distribution of the \red{$K$-band} absolute limiting magnitude $M_{\rm limit} = \mlim - DM$ (red crosses) and
absolute threshold magnitude $M_{\rm th} = \mth - DM$ (green circles) of the UKIDSS 
galaxies around each AGN as a function of AGN redshift.  
Where $DM$ is the distance modulus. 
$m_{\rm th}$ is a threshold magnitude defined in Equation (\ref{eq:det_eff}),
above which detection efficiency for galaxies becomes lower than 1.0. 
AGN samples are the same as Figure~\ref{fig:mlimit-z}. 
}
\label{fig:mlimit-abs-z}
\end{figure}

The limiting magnitude $\mlim$ of the UKIDSS galaxy samples is estimated for each AGN
sample, and the result is plotted in Figure~\ref{fig:mlimit-z}.
As can be seen from the figure, $\mlim$ distributes around 19.7 -- 20.3.
For each AGN, we also estimated threshold magnitude, $\mth$, 
below which detection efficiency for galaxies can be regarded as $100\%$. 
Figure~\ref{fig:mlimit-abs-z} shows the absolute magnitude corresponding to $\mlim$ and $\mth$ 
at AGN redshift. 
The detailed definition of the $\mlim$ and $\mth$ is described in the next section.

\begin{figure}
\plotone{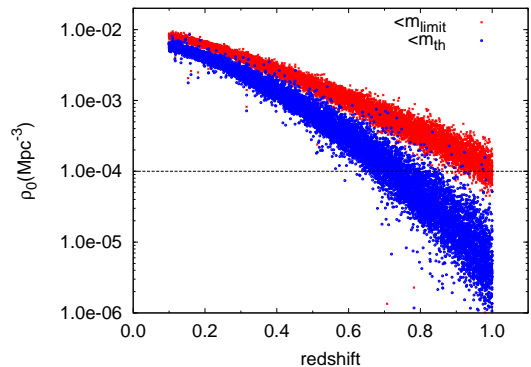}
\caption{Distribution of the average number density $\rho_{0}$ of detectable galaxies 
($<m_{\rm limit}$, red crosses) and galaxies brighter than $m_{\rm th}$ (blue circles) at the AGN redshift for each AGN sample.  
\red{Among the 11,335 AGN samples plotted in this figure, 
10,482 AGNs with $\rho_{0} > 10^{-4}{\rm Mpc}^{-3}$ for detectable galaxies (red crosses above the dashed line) are use in the following analysis. 
For the analysis of the complete galaxy sample in Section \ref{completeS}, we use 6,107 AGNs with $\rho_{0} > 10^{-4}{\rm Mpc}^{-3}$ for galaxies brighter than $m_{\rm th}$ (blue circles above the dashed line). } 
}
\label{fig:rho0_z}
\end{figure}

To remove the data of poor sensitivity, we adapted selection criteria of 
$\rho_{0} > 10^{-4}$ and $z < 1.0$, where $\rho_{0}$ is the average number density of galaxies 
detectable at the AGN redshift. 
$\rho_{0}$ was calculated by integrating the luminosity function upto the absolute 
magnitude corresponding to the apparent limiting magnitude $\mlim$ as described in the
next section and \citet{Shirasaki11}.
Figure~\ref{fig:rho0_z} shows the calculated $\rho_{0}$ as a function of AGN redshift.
In the same figure the number densities of complete sample (blue) are also shown, 
and they are calculated by integrating the luminosity function upto the absolute magnitude 
corresponding to $m_{\rm th}$.
The combined AGN catalog, which are based on catalogs of \citet{Shen11} and \citet{Greene07}, lists 32,806 AGNs at redshift between 0.1 to 1.0. 
11,335 objects are distributed in the survey area of UKIDSS LAS among them. 
In Figures \ref{fig:mlimit-z} -- \ref{fig:rho0_z}, values for areas within 1 degree around these 11,335 AGNs are plotted.  
10,482 AGNs are selected by the criterion of $\rho_{0} > 10^{-4}{\rm Mpc}^{-3}$ for $\mlim$. 

To reduce the effect of foreground clusters which are accidentally located near the sample AGNs on the sky, 
we rejected samples with anomalous distribution of surrounding galaxy by the method described in \citet{Shirasaki11}. 
\red{To reduce the effect of accidental alignment of the foreground cluster, we calculate the clustering coefficient, $B_{QG}$, around each AGN, which was defined as $\xi(r)=B_{QG} r^{-\gamma}$ \citep{Barr03}, where $\xi(r)$ is the cross-correlation function. 
$B_{QG}$ is calculated as 
$B_{QG}=\frac{3-\gamma}{2\pi C(\gamma)}\frac{N_{\rm total}-N_{\rm bg}}{\rho_0}(1{\rm Mpc})^{\gamma-3}$
, where $N_{\rm total}$ is the total number of observed galaxies at $r_p <1$ Mpc, where $r_p$ is a perpendicular distance, and $N_{\rm bg}$ is the expected background count at $r_p < 1$Mpc (see Section \ref{analyS} for the definition of $\xi$, $r_p$, $\gamma$, and $C(\gamma)$). 
We reject AGN samples with $|B_{QG}|>10^4$ (30 times the clustering coefficient of the Abell class 0 objects).
In addition, we select sample AGNs without the effect from the nearby cluster located in regions offset from the AGNs. 
We count number density $n(r_{\rm p})$ of UKIDSS galaxies for each circular region with $\Delta r_p=0.2$Mpc width around AGNs, and compute their statistical error. 
We adopt the following criteria: 
reduced $\chi^2$ of the radial number density relative to the flat distribution is $\chi^2/(n-1)\leq 3$; 
the maximum deviation, $\sigma_{\rm max}$, of the $n(r_{\rm p})$ is smaller than $5\sigma$. 
In Section~\ref{resultS}, we also show the results without these selection for comparison. }

\begin{figure}
\plotone{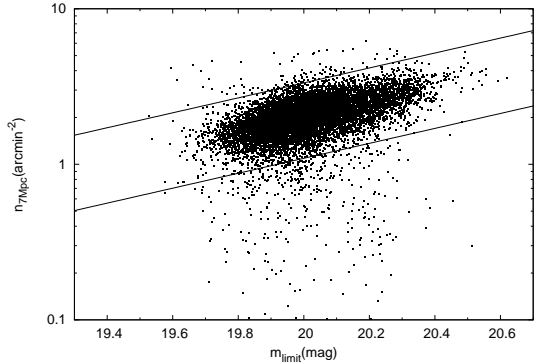}
\caption{Distribution of the surface number density of galaxies around AGNs as a function of the \red{$K$-band} limiting magnitude. 
$n_{7 \rm Mpc}$ is the surface density of UKIDSS galaxies in the circular area around the AGN within an angular radius corresponding to 7Mpc(comoving) at the AGN redshift. 
We select AGN samples whose deviations of $\log n_{7 \rm Mpc}$ are less than $1.5\sigma$ (between the two solid lines) to reject effect of foreground clusters and bad quality regions of UKIDSS data. 
AGN samples are the same as Figure~\ref{fig:mlimit-z}. 
}\label{mlim-n}
\end{figure}

We also present result of analysis for the AGN sub-sample whose deviations of surface number density of UKIDSS sources is smaller than $1.5\sigma$ at limiting magnitude $\mlim$ to check again the effect of foreground objects. 
Figure~\ref{mlim-n} shows the projected number density, $n_{7\rm Mpc}$, of galaxies in the area around the AGN with an angular radius corresponding to 7Mpc (comiving) at the AGN redshit as a function of $\mlim$. 
Because the projected number density of the foreground or background galaxies is $\sim 30$ times larger than the galaxies in the host clusters even for the highly clustered region with $r_0 \sim 10$Mpc, this criterion rarely rejects sample AGNs with the real overdensities around them, but rejects AGNs located near the foreground clusters.  

\begin{figure}
\plotone{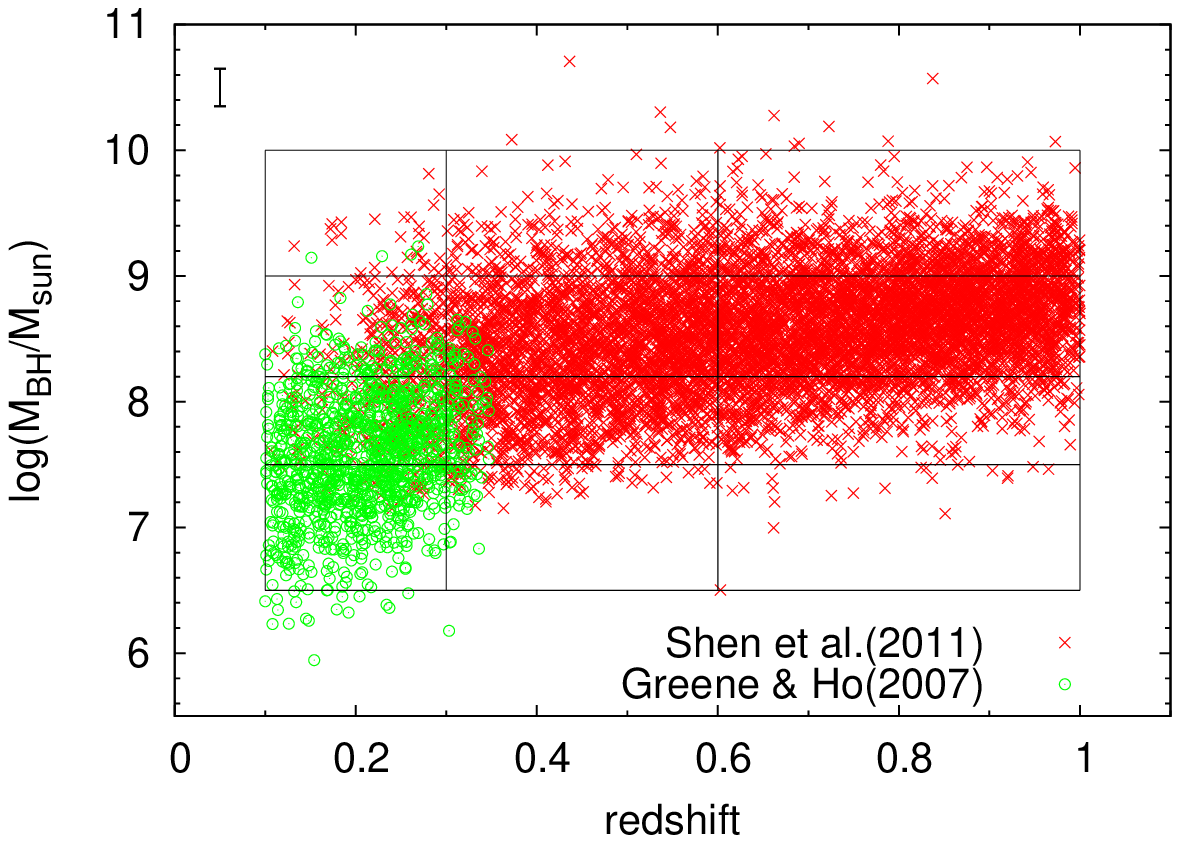}
\caption{Distribution of 9,394 AGNs used in the following analysis in mass-redshift space.
Red crosses and green circles denote samples by \citet{Shen11} and \citet{Greene07}, respectively. 
For the samples by \citet{Greene07}, we plotted the recomputed virial mass by means of the parameter values of the mass estimator in \citet{Shen11}. 
The superposed grid indicates the sub-samples which we used to explore mass and redshift dependences. 
Bar at top left corner shows typical error for virial BH mass estimation. 
}
\label{sample}
\end{figure}

We used 9,394 AGNs, which were selected by criteria described in \citet{Shirasaki11}, in the following analysis. 
8,060 AGNs were selected by the $1.5\sigma$ criterion for $n_{7\rm Mpc}$. 
Figure~\ref{sample} shows the distribution of the virial mass of the 9,394 SMBHs as a function of the redshift. 
1,202 AGNs were extracted from the catalog of \citet{Greene07} (red crosses) and the others were from the catalog of \citet{Shen11} (green circles).
For the samples by \citet{Greene07}, we plot the recomputed virial mass using the parameters of mass estimator in \citet{Shen11}. 
3,749 objects among the samples from \citet{Shen11} had been identified by the uniform criteria of SDSS quasars \citep{Richards02}. 
We also present the results of the clustering analysis for these sub-samples. 
We divided our sample into four mass bins of $\log(\Mbh / \msun) =$ 6.5--7.2, 7.2--8.0, 8.0--9.0, and $9.0$--$10.0$, and three redshift bins of $z=$0.1--0.3, 0.3--0.6, 0.6--1.0, to see mass and redshift dependences. 
The number, the averaged mass and the averaged redshift of AGN samples for each mass range and redshift range is listed in Table~\ref{Tresult}. 

For $\sim 6\%$ of AGNs in our sample, radio counterparts were found in the FIRST source catalog \citep{Schneider10}. 
The percentage of objects with radio counterparts is higher for AGNs with higher BH mass. 
The fraction of AGNs with radio counterpart is $14\%$ at $\Mbh>10^9\msun$. 
For $\sim 12\%$ of the sample AGNs, X-ray counterparts were found in the ROSAT catalog. 
The fraction of AGNs with X-ray detection is almost the same for all the BH mass ranges.

\section{Analysis Method}\label{analyS}
We have followed the method described in \citet{Shirasaki11} for the cross-correlation analysis between AGNs and galaxies. 

The cross-correlation function of AGNs and galaxies $\xi(r)$ can be expressed as an 
excess of number density of galaxies $\rho(r)$ relative to the average density $\rho_{0}$ 
at the AGN redshift.
\begin{equation}
\xi(r) = \frac{\rho(r)}{\rho_{0}} - 1,
\end{equation}
where $r$ represents the distance from an AGN.

In this analysis the redshift of galaxies are not measured, thus a projected cross 
correlation function $\omega(r_{p})$ is calculated from projected number densities
of galaxies $n(r_{p})$:
\begin{equation}
\omega(r_{p}) = \frac{n(r_{p}) - n_{\mbox{\scriptsize bg}}}{\rho_{0}},
\label{eq:omega_1}
\end{equation}
where $r_{p}$ represents projected distance from an AGN at the redshift, and
$n_{\mbox{\scriptsize bg}}$ represents the surface density of background/foreground
galaxies.

\subsection{Estimating the Average Density and the Limiting Magnitude}\label{rho0S}
Estimation of $\rho_0$ is crucial for this analysis. However, we cannot estimate $\rho_0$ 
directly from the data itself since the information on the redshift is lacking.
In this work, we estimated $\rho_0$ based on the luminosity function of galaxies
obtained by previous studies \citep{Cirasuolo07,Gabasch04,Gabasch06,Kochanek01,Montero-Dorta09}.
The luminosity function $\phi(M;z,\lambda)$ is parametrized as a function of 
redshift $z$ and rest-frame wavelength $\lambda$ as follows. 

\begin{figure}
\includegraphics[width=0.7\columnwidth]{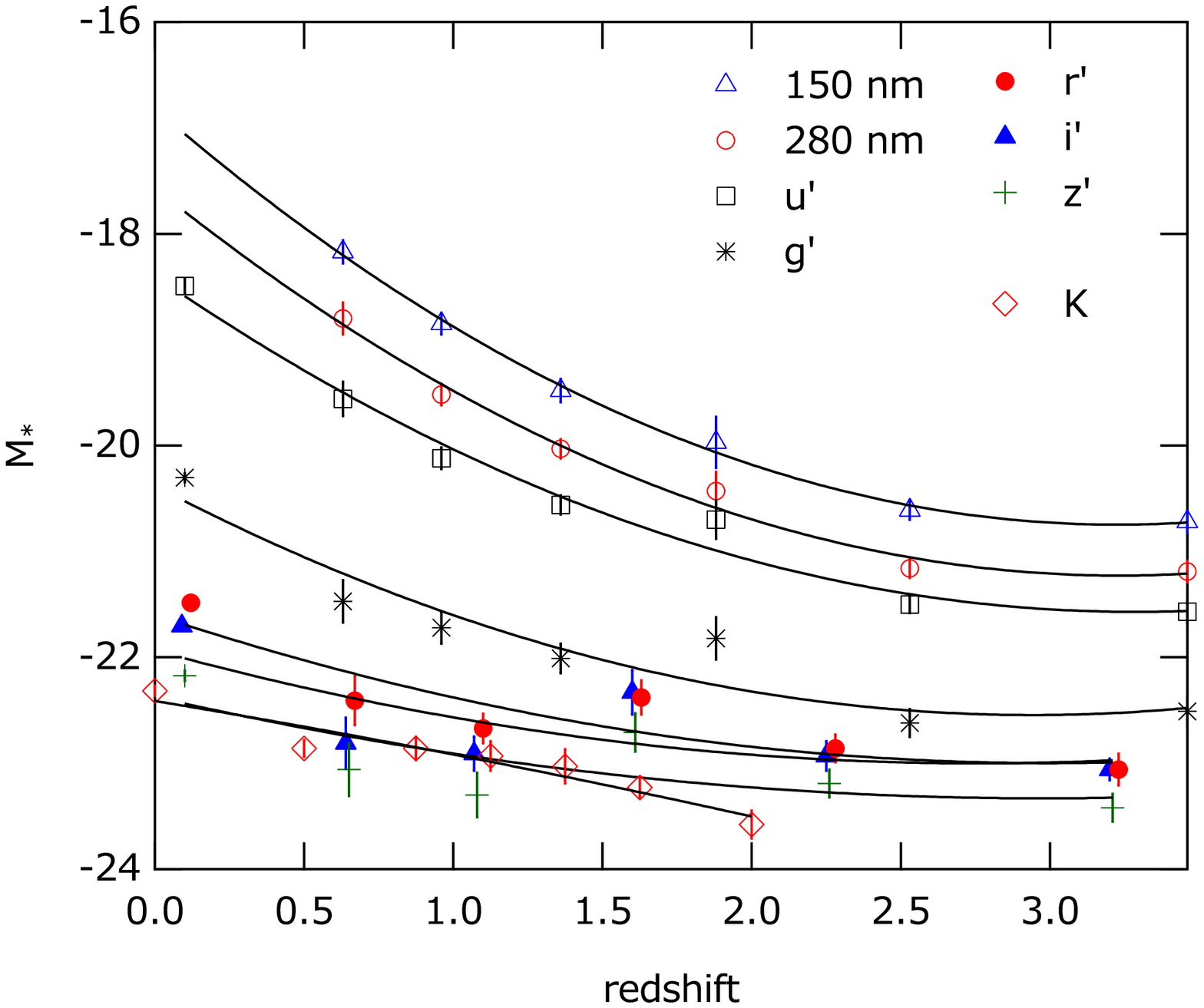}
\includegraphics[width=0.7\columnwidth]{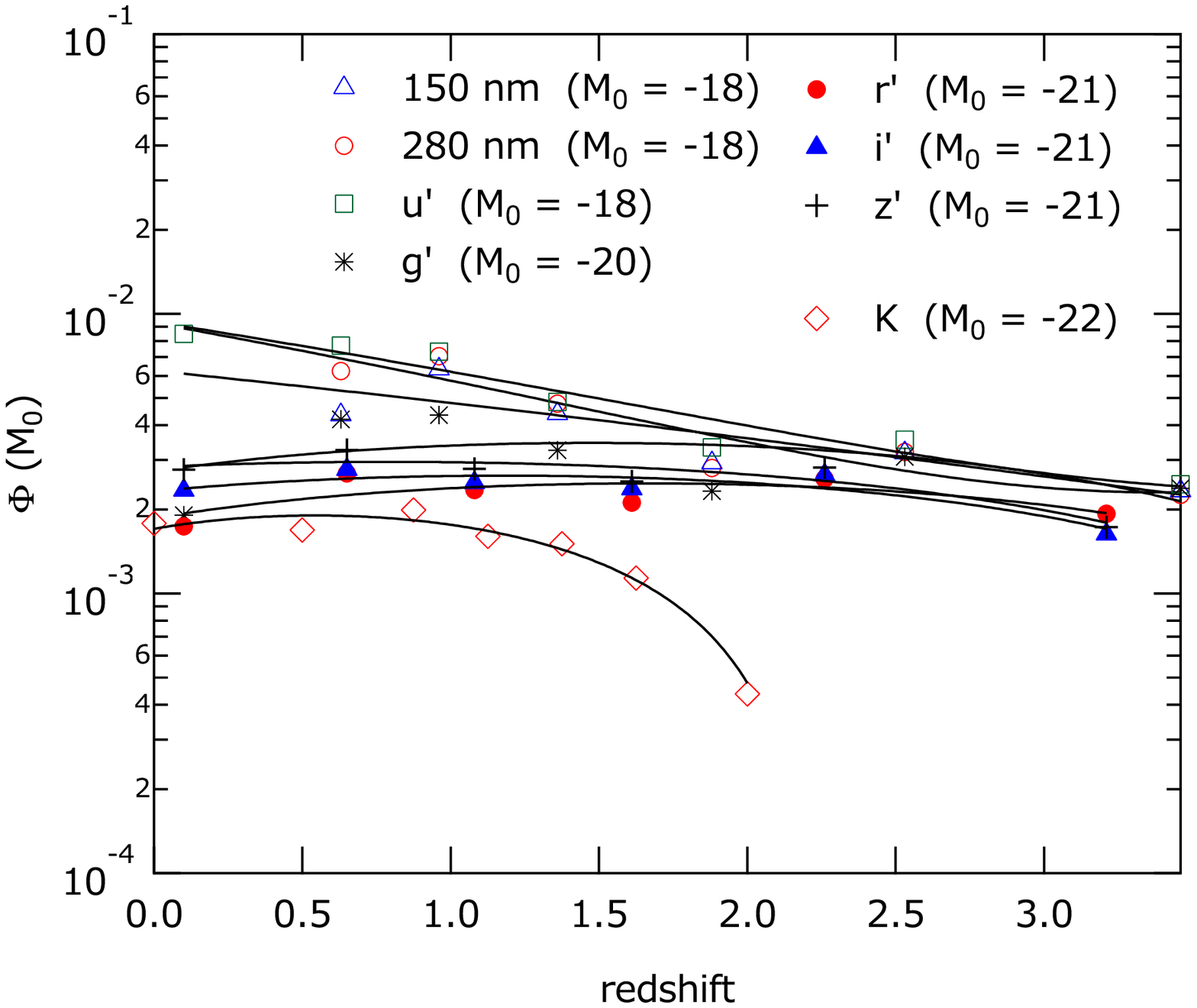}
\caption{Parameters of the Schechter function derived by \red{
Cirasuolo et al. 2007 ($K$-band. $z\geq0.5$), 
Gabasch et al. 2004 (150nm, 280nm, $u'$, and $g'$. $z\geq0.6$)
Gabasch et al. 2006 ($r'$, $i'$, and $z'$. $z\geq0.6$)
Kochanek et al. 2001 ($K$-band. $z=0$)
Montero-Dorta $\&$ Prada 2009 ($u'$, $g'$, $r'$, $i'$, and $z'$. $z=0.1$). }
The solid lines represent fitting functions to
parametrize $M_{*}$ and $\Phi(M_{0})$ as a function redshift.
{\it Top panel}: $M_{*}$ for each rest-frame wavelength band. 
{\it Bottom panel}: Number densities $\Phi(M_{0})$ at a reference magnitude $M_{0}$.
The reference magnitude $M_{0}$ for each wavelength band is $-18$ for 150~nm,
280~nm, and $u'$ band, $-20$ for $g'$ band, $-21$ for $r'$, $i'$, and $z'$ band, 
and $-22$ for $K$ band. 
}\label{fig:Mstar}
\end{figure}

We use the Schechter function to represent the 
luminosity function,
\begin{equation}
\phi(M) = 0.4 \cdot \phi_{*} \cdot \ln(10) \cdot 10^{-0.4 (M - M_{*}) (\alpha + 1)}
              \cdot \exp{[-10^{-0.4 (M - M_{*})}]}.
\label{eq:Schechter}
\end{equation}
In Figure~\ref{fig:Mstar}, 
$M_{*}$ and $\phi(M_{0})$ derived from the fitting function 
obtained in the works by
\citet{Cirasuolo07,Gabasch04,Gabasch06,Kochanek01,Montero-Dorta09}
are plotted as a function of redshift for each rest-frame wavelength.
$M_{0}$ is a reference magnitude where the luminostiy function is normalized
and parametrized as a function of redshift, and it is selected at a dimmer
side of the $M_{*}$ magnitude.
Those data points are fitted with a 3rd degree polynomial function of redshift
as shown in the Figure~\ref{fig:Mstar} as solid lines.
The standard deviations of $M_{*}$ and $\phi(M_{0})$ from the fitted functions are
0.2 mag and 15\%, respectively.
For the parameter $\alpha$, we used fixed values such as 
$-$1.1 for $\lambda < 400$nm, 
$-$1.25 for $400 \le \lambda < 1000$nm, and
$-$1.0 for $\lambda \ge 1000$nm.

As $M_{*}$ and $\phi(M_{0})$ at an arbitrary redshift for the eight wavelength bands
can be calculated using the parametrization derived above, these parameters
at an arbitrary wavelength are derived by interpolating them as a function
of wavelength with a cubic spline method.
In this way, we parametrized the luminosity function as function of redshift and
rest-frame wavelength.
\red{We estimate $\rho_0$ by means of the Schechter function with the parameters at the AGN redshift and wavelength corresponding to the observed $K$ band. }

$\rho_0$ can be calculated by integrating $\phi(M;z,\lambda)$ to the absolute magnitude $M$ 
at an AGN redshift corresponding to an apparent limiting magnitude $\mlim$, 
\begin{equation}
\rho_0 = \int^{\mlim-DM(z)}_{m_{\rm low}-DM(z)} \phi(M;z,\lambda) dM, 
\label{eq:rho0_1}
\end{equation}
where $DM$ is the distance modulus and $m_{\rm low}$ is a lower boundary of the apparent 
magnitude. 

As the limiting magnitude varies among the AGN samples, it was estimated from the
measured magnitude distribution $N(m)$ as explained below.
The observed magnitude distribution $N_{\rm obs}(m)$ can be expressed as a multiplication
of the {\it true} magnitude distribution $N_{\rm true}(m)$ and the detection efficiency
$DE(m)$:
\begin{equation}
N_{\rm obs}(m) = N_{\rm true}(m) \times DE(m).
\end{equation}
We model $N_{\rm true}(m)$ and $DE(m)$ as introduced in \citet{Shirasaki11}:
\begin{eqnarray}
   N_{\rm true}(m)  = \left\{
   \begin{array}{ll}
     c \cdot 10^{a (m-m_{\rm b})}    & ( m < m_{\rm b}   )  \\
     c \cdot 10^{b (m-m_{\rm b})}    & ( m \ge m_{\rm b} ), \\
   \end{array}
   \right.
   \label{eq:mag_dist_org}
\end{eqnarray}
\begin{eqnarray}
   DE(m)  = \left\{
   \begin{array}{ll}
     1  & ( m < m_{\rm th} )  \\
     \exp{( -(m-m_{\rm th})^2/\sigma_{m}^{2} ) }  & ( m \ge m_{\rm th}), \\
   \end{array}
   \right.
   \label{eq:det_eff}
\end{eqnarray}
By fitting the model function of $N_{\rm obs}(m)$ to the observed magnitude distribution,
we obtained the model parameters $a$, $b$, $c$, $m_{\rm b}$, $m_{\rm th}$ and $\sigma_{m}$
for each area around an AGN sample.
We determine $\rho_{0}$ as:
\begin{equation}
\rho_{0} = \int^{\infty}_{m_{\rm low} - DM}  \phi(M; z, \lambda) DE(M+DM) dM
\label{eq:rho0_2}
\end{equation}
Equating the right-hand sides of Equation~(\ref{eq:rho0_1}) and (\ref{eq:rho0_2}),
$\mlim$ is derived.                   

\begin{figure}
\plotone{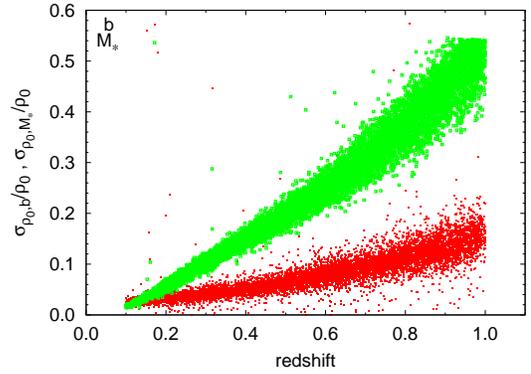}
\caption{Fraction of uncertainty of $\rho_{0}$ originated from the uncertainty of the
model parameter $b$ (red filled square) and $M_{*}$ (green open square) of the {\it true} magnitude distribution. 
\label{fig:ErrorRho0}
}
\end{figure}

The uncertainty of $\rho_0$ determined as explained above is dominated with the 
uncertainties of the model parameters $b$ and 
uncertainties of parametrization of the luminosity function, and they are taken into
account as a systematic error in estimating the cross correlation length.
We estimate the uncertainty of $b$ as 0.04, which comes from the standard 
deviation of $b$ calculated for each AGN sample.
Considering the uncertainty of $b$, corresponding uncertainties of $m_{\rm th}$ and
$\sigma_{m}$ are estimated by comparing these fitting parameters obtained
by fixing the $b$ parameter to $b_{\rm best} \pm 0.04$, where $b_{\rm best}$ is the 
best fitting parameter for the AGN sample.
The uncertainty of $\rho_{0}$ originated from the 
uncertainties of $m_{\rm th}$ and $\sigma_{m}$ (i.e. $b$) are calculated with error 
propagation. 
The uncertainly $\sigma_{\rho_{0},b}$ are plotted as a function of redshift in Figure~\ref{fig:ErrorRho0}.

To evaluate the uncertainty of $\rho_{0}$ due to the uncertainties of 
parametrization of the luminosity function, we assumed uncertainties of
$M_{*}$ and $\phi(M_{0})$ to be 0.2 mag and 15\% respectively, which are 
the standard deviation of data points in Figure~\ref{fig:Mstar} from the fitting functions.

$\sigma_{\rho_{0},M_{*}}$ is the uncertainty 
of $\rho_{0}$ due to the uncertainty of $M_{*}$, and  
is plotted in Figure~\ref{fig:ErrorRho0}. 
The uncertainty of $\rho_{0}$ due to the uncertainty of $\phi(M_{0})$ is independent from redshift and is constant
value of $\sigma_{\rho_{0},\phi(M_{0})}=$0.15.
Then the total uncertainty of $\rho_{0}$ is calculated as:

\begin{equation} \label{Eq:sigma_rho0}
\sigma_{\rho_{0}}^{2} = \sigma_{\rho_{0},b}^{2} + \sigma_{\rho_{0},M_{*}}^{2} + \sigma_{\rho_{0},\phi(M_{0})}^{2}
\end{equation}

\subsection{Cross Correlation}
We assume the power-law form for the cross-correlation function, 
\begin{equation}
\xi(r) = \left( \frac{r}{r_{0}} \right) ^{-\gamma}, 
\end{equation}
where $r_0$ is a correlation length, and $\gamma$ is a power-law index and 
fixed to 1.8, which is a canonical value measured by many other works.
Then the projected cross-correlation function can be expressed as:
\begin{eqnarray}
\omega(r_p) = 2 \int^\infty_0 \xi(r_p,\pi) d\pi 
            = 2 \int^\infty_{r_p} \frac{r \xi(r)}{\sqrt{r^2-r_p^2}} dr \nonumber \\ 
            = r_p \left( \frac{r_0}{r_p} \right) ^\gamma 
              \frac{\Gamma(\frac{1}{2})\Gamma(\frac{\gamma-1}{2})}{\Gamma(\frac{\gamma}{2})}, 
\label{eq:omega_2}
\end{eqnarray}
where $\pi$ and $r_p$ are distance along and perpendicular to the line of sight, respectively, 
and $\Gamma$ is the Gamma function. 

From equations(\ref{eq:omega_1}) and (\ref{eq:omega_2}), the projected number density 
of galaxies around an AGN can be modeled as:
\begin{equation}
n(r_{p}) = C(\gamma) \times \rho_{0} \times  r_{p} \left( \frac{r_{0}}{r_{p}} \right)^{\gamma} 
         + n_{\mbox{\scriptsize bg}},
\end{equation}
where the term of the Gamma function is represented by $C(\gamma)$.
By fitting this model function to the observed projected number density, we can 
derive the model parameters $r_{0}$ and $n_{\rm bg}$. 
$\rho_0$ is determined by the method described in Section~\ref{rho0S}
Since the clustering signature for each AGN is too weak to derive the parameters 
individually, we applied this fitting to the averages of $n(r_p)$ and $\rho_0$ for a given AGN group,
\begin{equation}\label{eq:n-rtotal}
\langle n(r_{p})\rangle = C(\gamma) \times \langle \rho_{0} \rangle \times  r_{p} \left( \frac{r_{0}}{r_{p}} \right)^{\gamma} 
         + \langle n_{\mbox{\scriptsize bg}} \rangle,
\end{equation}
and derive $r_{0}$ and $\langle n_{\rm bg} \rangle$. 
The uncertainty of $r_{0}$ is calculated as square root of square sum of the systematic error derived from
the uncertainty of $\rho_0$ described in equation~(\ref{Eq:sigma_rho0}) and statistical error of 1$\sigma$ 
by fitting $\langle n(r_p) \rangle$. 
It should be noted that the cross-correlation function obtained by this method
is not an average for the AGN group, but an average weighted with $\rho_{0}$.
Thus the result is biased to the low-z and high-sensitivity samples.

\section{Results}\label{resultS}
\begin{figure*}

\begin{tabular}{cc}

\begin{minipage}{0.5\hsize}
  \begin{center}
\plotone{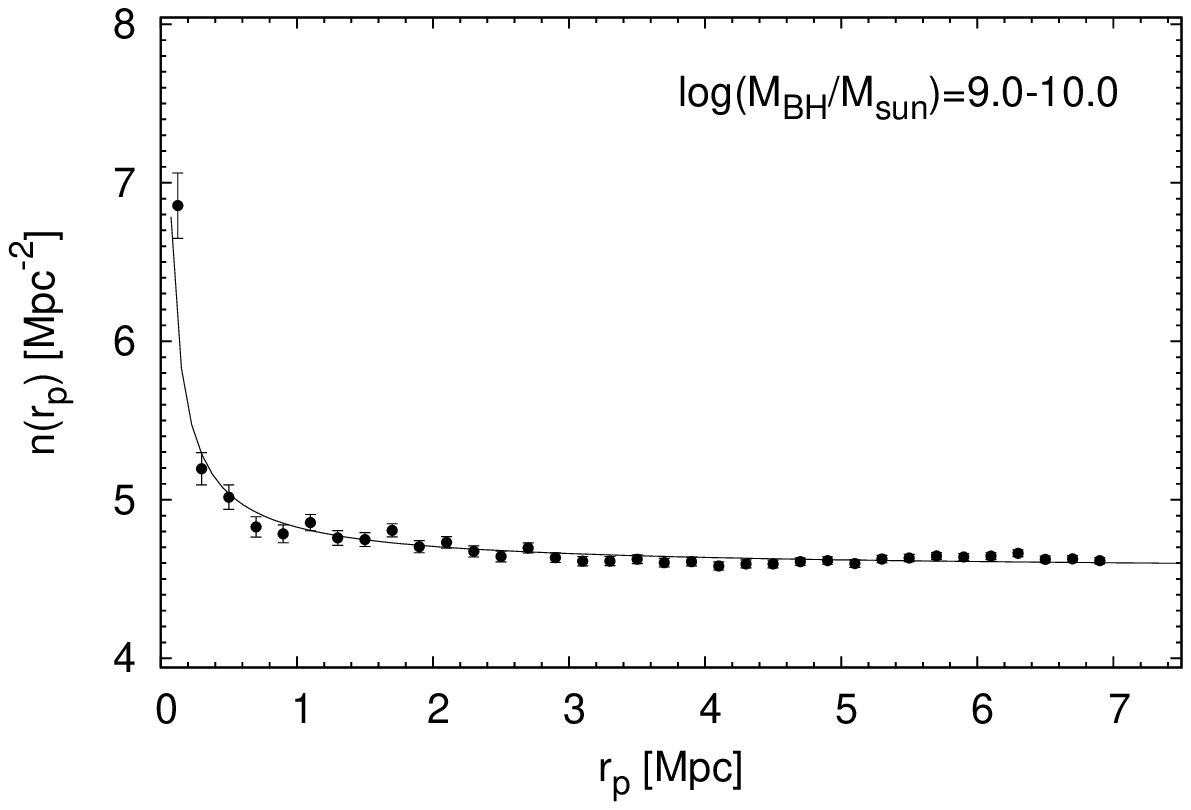}
\end{center}
\end{minipage}

\begin{minipage}{0.5\hsize}
  \begin{center}
\plotone{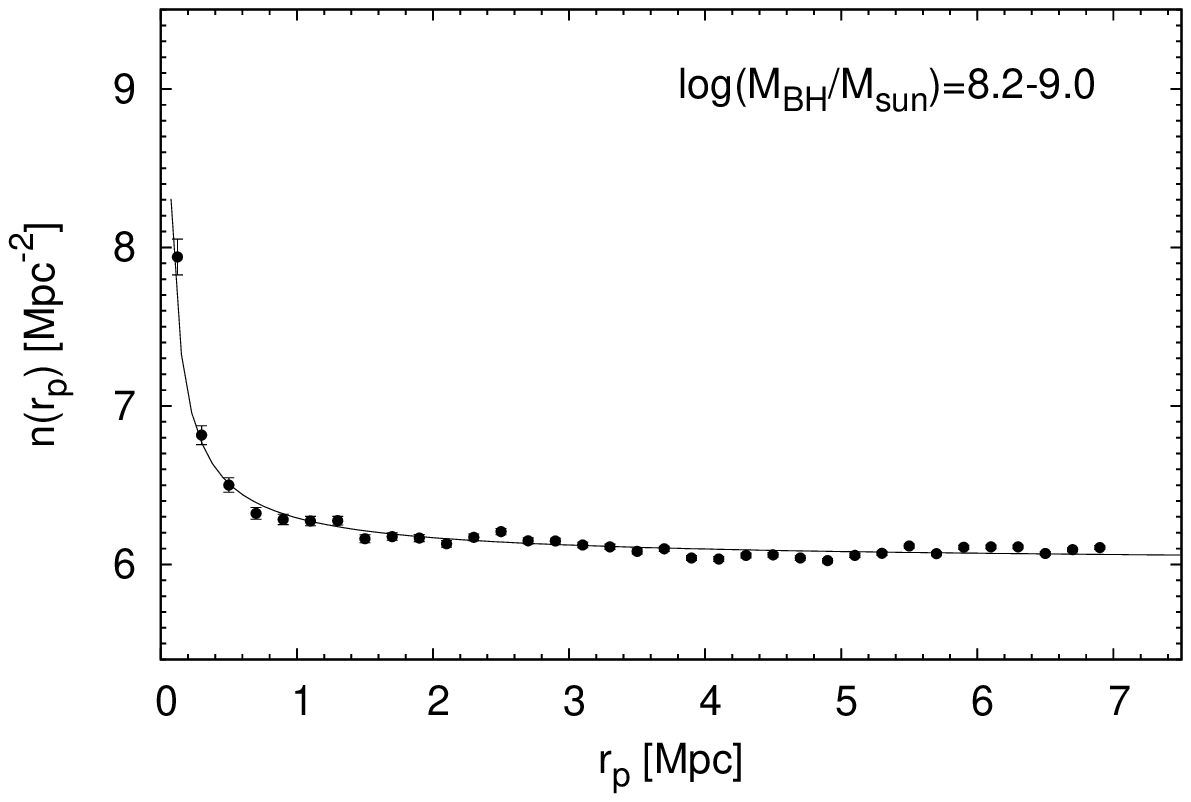}
\end{center}
\end{minipage}\\

\begin{minipage}{0.5\hsize}
  \begin{center}
\plotone{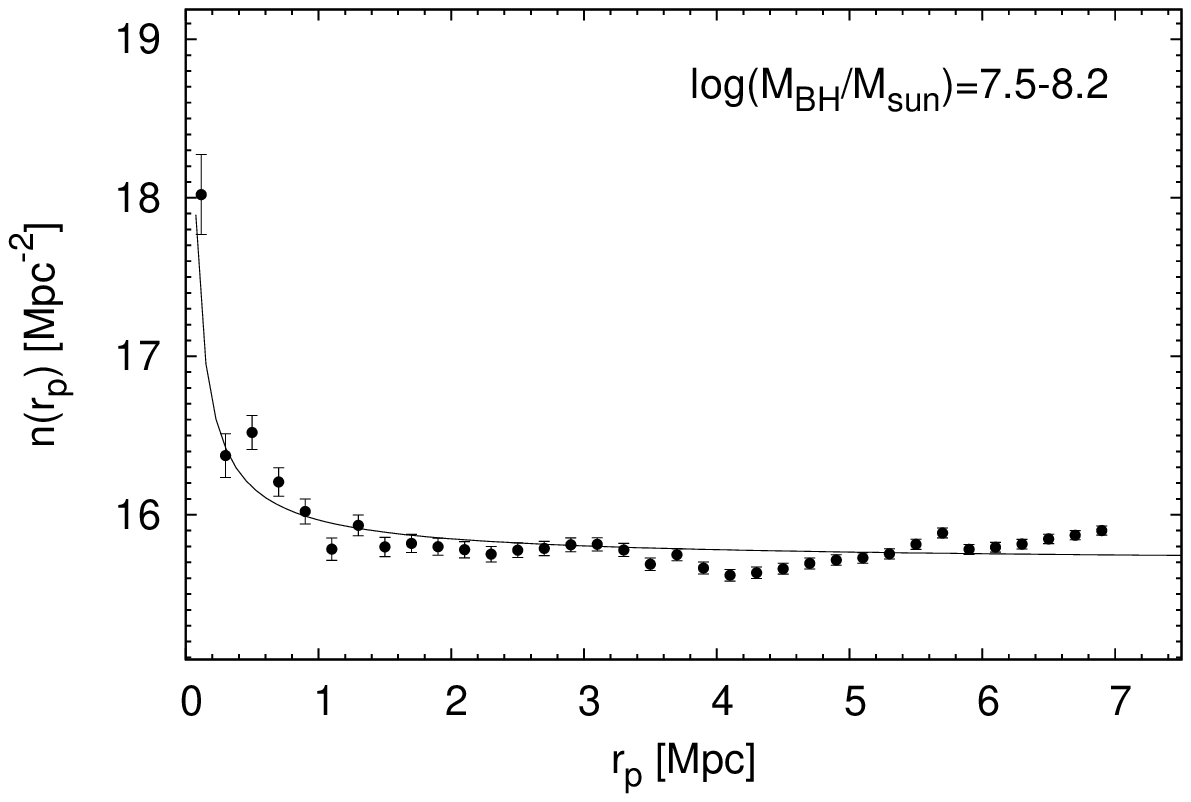}
\end{center}
\end{minipage}

\begin{minipage}{0.5\hsize}
  \begin{center}
\plotone{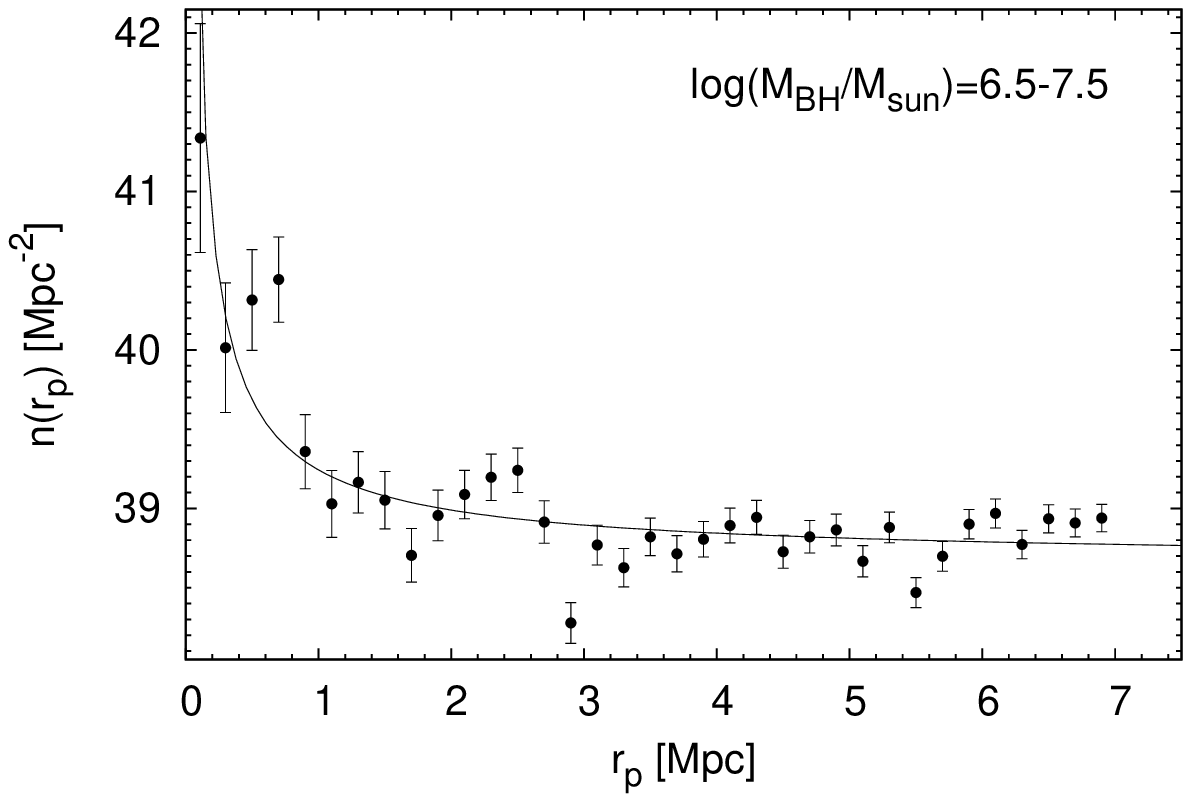}
\end{center}
\end{minipage}

\end{tabular}
\caption{Projected number density of UKIDSS sources against projected distance from an AGN. 
Four panels show results for AGN samples in the four different mass ranges: 
$\log(\Mbh/\msun)=9.0$ -- $10.0$ (top left), $8.2$ -- $9.0$ (top right), $7.5$ -- $8.2$ (bottom left), and $6.5$ -- $7.5$ (bottom right). 
Poisson error bars of projected number density are attached. 
The solid lines denote least-$\chi^2$ fit by the power function (Eq.~\ref{eq:n-rtotal}). 
}
\label{n-r}
\end{figure*}

\begin{figure*}

\begin{tabular}{cc}

\begin{minipage}{0.5\hsize}
  \begin{center}
\plotone{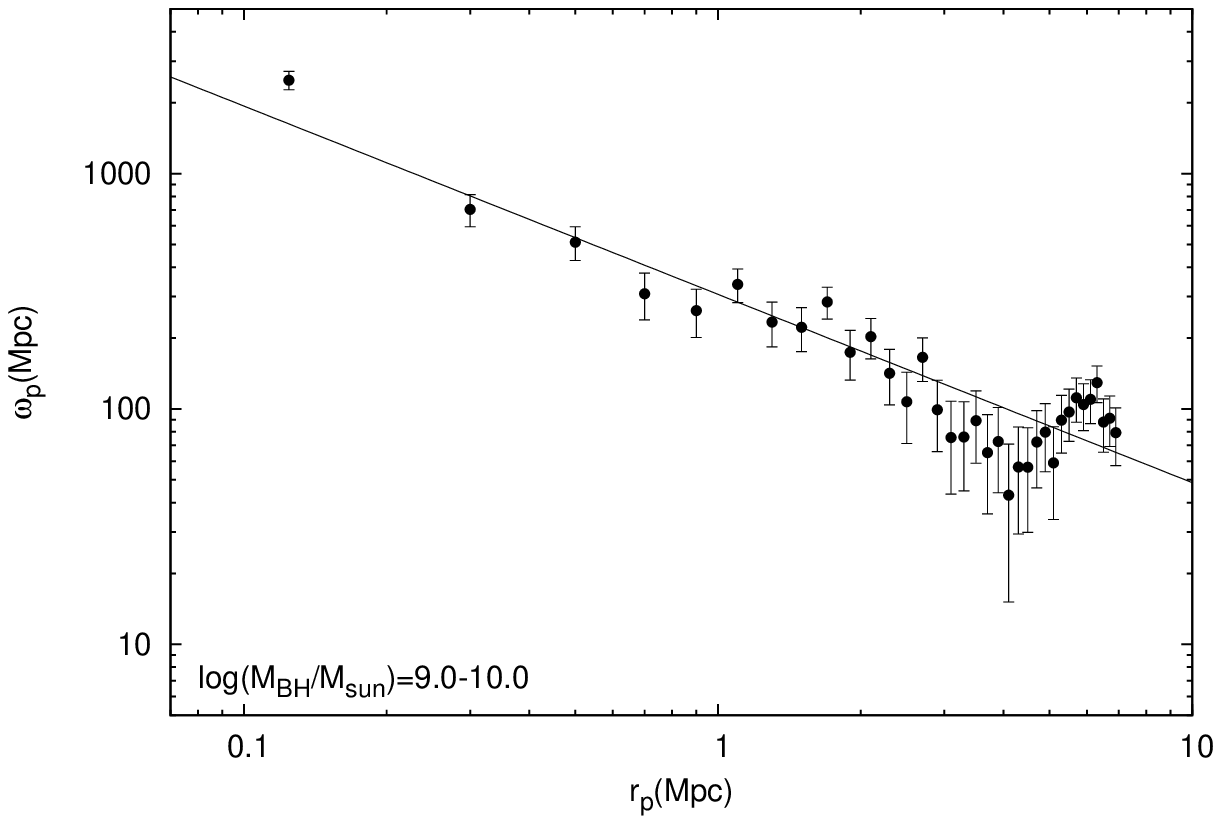}
\end{center}
\end{minipage}

\begin{minipage}{0.5\hsize}
  \begin{center}
\plotone{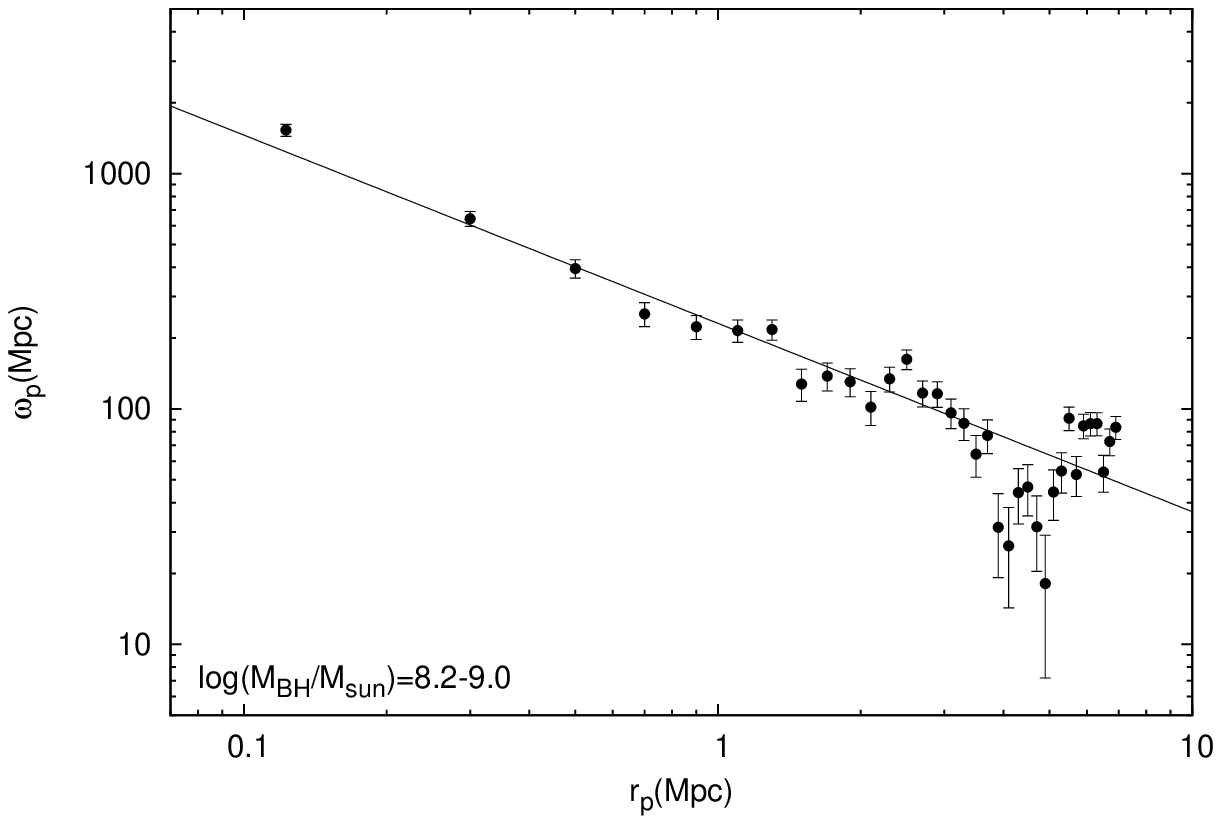}
\end{center}
\end{minipage}\\

\begin{minipage}{0.5\hsize}
  \begin{center}
\plotone{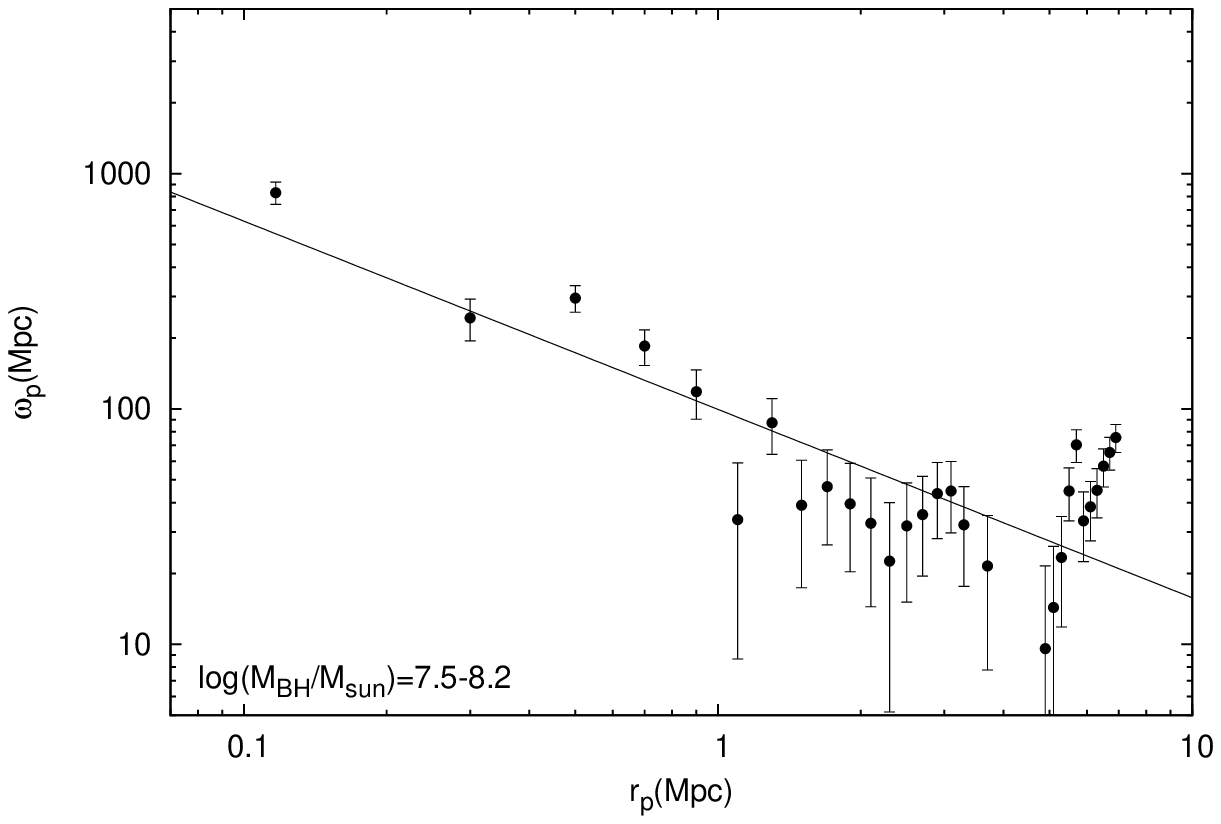}
\end{center}
\end{minipage}

\begin{minipage}{0.5\hsize}
  \begin{center}
\plotone{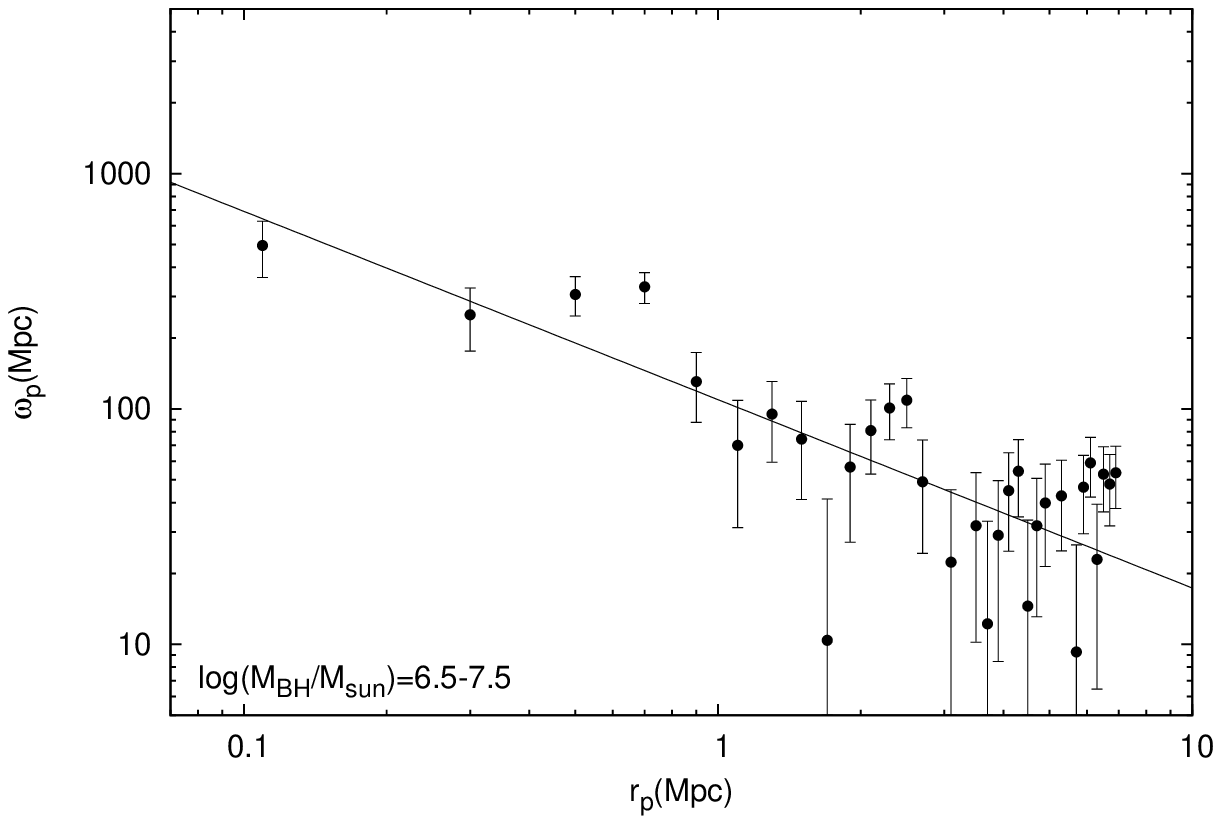}
\end{center}
\end{minipage}

\end{tabular}

\caption{Projected cross correlation function of UKIDSS sources against projected distance from an AGN. 
Four panels show results for AGN samples in the four different mass ranges. 
Error bars show the uncertainty due to statistical error of projected number density. 
The solid lines denote least-$\chi^2$ fit by the power function (Eq.~\ref{eq:omega_2}). 
}
\label{rho-r}
\end{figure*}

By the analysis described above, we have estimated the scale length of AGN-galaxy cross correlation for the whole sample to be $r_0= 5.8^{+0.8}_{-0.6} \hMpc$. 
This is comparable with or slightly smaller than results of the previous studies of AGN-galaxy cross-correlation ($r_0=5.95\pm0.90 \hMpc$ for X-ray AGNs at $z=0.7$--1.4, \citep{Coil09}; $6.98\pm0.6 \hMpc$ for optical AGNs at $z<1$, \citep{Mountrichas09}; $r_0=6.0\pm0.5 \hMpc$ for optical AGNs at $0.2<z<0.6$, \citep{Padmanabhan09}).  
\citet{Donoso10} found $r_0=8.35\pm0.09$ for radio AGNs and $r_0=5.02\pm0.24 \hMpc$ for optical AGNs at $z=0.4$--0.8.
\citet{Krumpe12} derived that $r_0=6.91^{+0.17}_{-0.18}\hMpc$ at $z=0.16$--0.36 between SDSS AGNs and the SDSS main-galaxy sample and $r_0=7.21^{+0.21}_{-0.22}\hMpc$ at $z=0.36$--$0.5$ for SDSS AGNs and luminous red galaxies. 

Now, we present the results of cross-correlation analysis adopted for the four mass ranges described in Figure~\ref{sample} and Table~\ref{Tresult}, to see the dependence of the clustering amplitude on BH mass. 
Figures~\ref{n-r} and \ref{rho-r} show the measured projected number density $n(r_p)$ and projected cross-correlation function $\omega(r_p)$, respectively, for each mass range. 
Four panels represent results for the mass ranges of $\log(\Mbh/\msun)=9.0$ -- $10.0$ (top left),  $8.2$ -- $9.0$(top right), $7.5$ -- $8.2$ (bottom left), and $6.5$ -- $7.5$ (bottom right). 
The projected number density of the areas of each circular ring is plotted with the Poisson error bars.
Solid lines denote least-$\chi^2$ fitting by the power-law (Eq.~\ref{eq:n-rtotal}). 
The fitting parameters are summarized in Table~\ref{Tresult}. 

\begin{figure}
\plotone{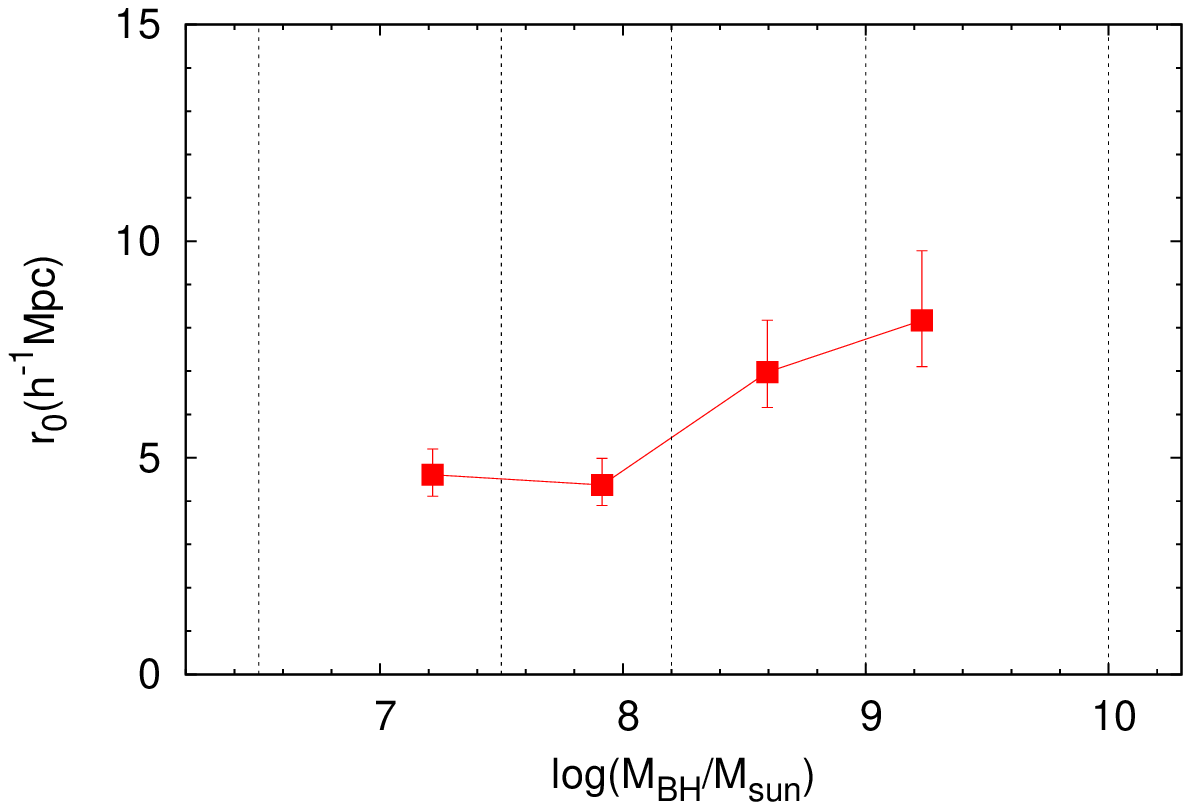}
\caption{The scale length, $r_0$, of the cross-correlation between AGNs and galaxies as a 
function of virial mass of SMBH. 
Error bars of $r_0$ denote square root of square sum of the systematic error derived from
the uncertainty of the estimation of $\rho_0$ and statistical error of 1$\sigma$. 
Vertical dotted lines show boundaries of mass ranges.  
}\label{m-r0}
\end{figure}

Figure~\ref{m-r0} shows the scale length $r_0$ of cross-correlation as a function of virial mass of SMBHs. 
Error bars of $r_0$ denote square root of square sum of the systematic error calculated from $\sigma_{\rho_0}$ described in Section~\ref{rho0S} and the statistical error of 1$\sigma$. 
We can see a trend that $r_0$ increases as mass increases for $\Mbh>10^8\msun$. 
When we consider only the statistical error, the significance of difference of $r_0$ is $9.6\sigma$ for between $\log(\Mbh/\msun)=7.5$ -- $8.2$ bin and $8.2$--$9.0$ bin, and $3.4\sigma$ for between $8.2$--$9.0$ bin and $9.0$--$10.0$ bin. 
For $\Mbh\lesssim10^8\msun$, we cannot see the significant mass dependence. 
These trends are also seen for the dataset grouped with finer mass ranges. 

\begin{figure}
\plotone{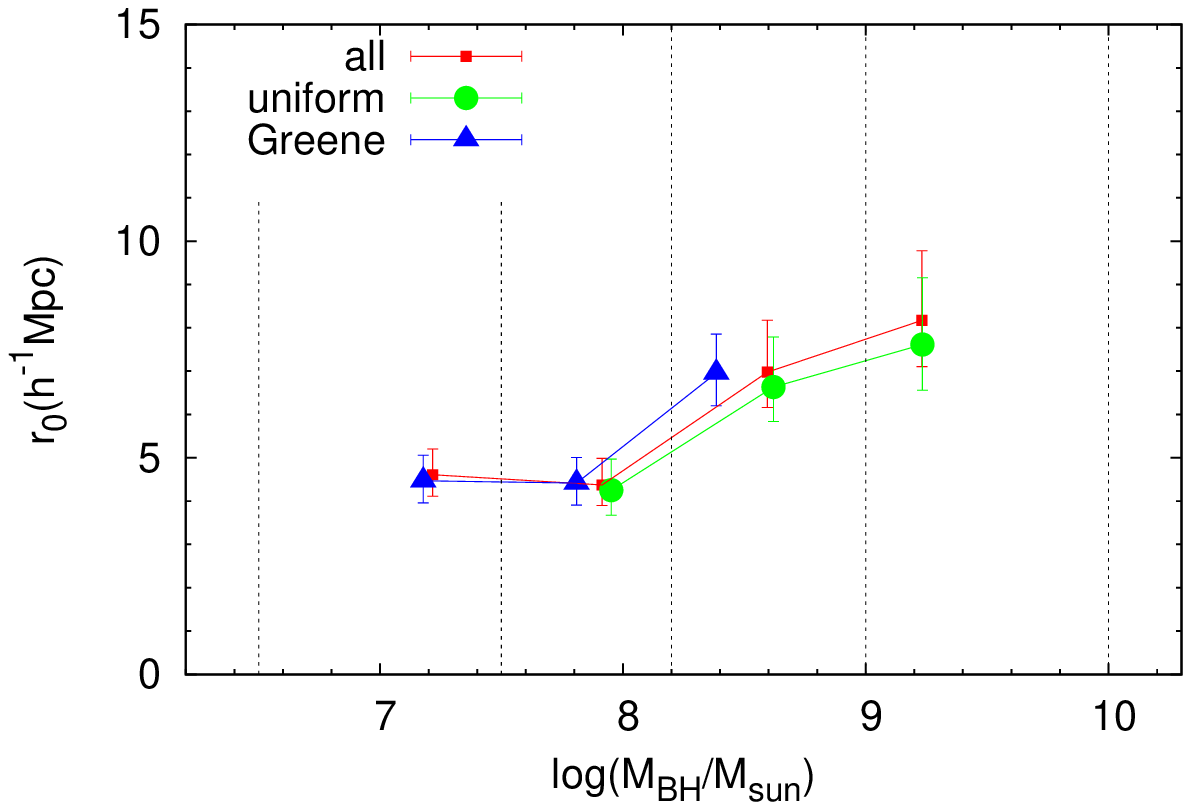}
\caption{Comparison between source catalogs. 
Red squares are the results for the total sample (the same with Figure~\ref{m-r0}). 
Green circles show the results for AGNs which are selected by uniform criteria of SDSS \citep{Richards02} among the samples from the catalog of \citet{Shen11}. 
Blue triangles show the results for the AGN samples extracted from the catalog of \citet{Greene07}. 
}
\label{unif-greene}
\end{figure}

Figure~\ref{unif-greene} shows $r_0$ for AGN samples from the catalog of Shen et al. 2011 (green circles) which are identified by the uniform criteria of SDSS \citep{Richards02} and for samples of Greene \& Ho 2007 (blue triangles). 
As seen in the figure, the mass dependences of both sub-samples are quite similar. 
We can see, however, a little offset between the results for the two sub-samples. 
This can be due to the systematic difference of the BH mass estimate with $\sim0.2$ dex between the two catalogs as mentioned in Section~\ref{dataS}. 

\begin{figure}
\plotone{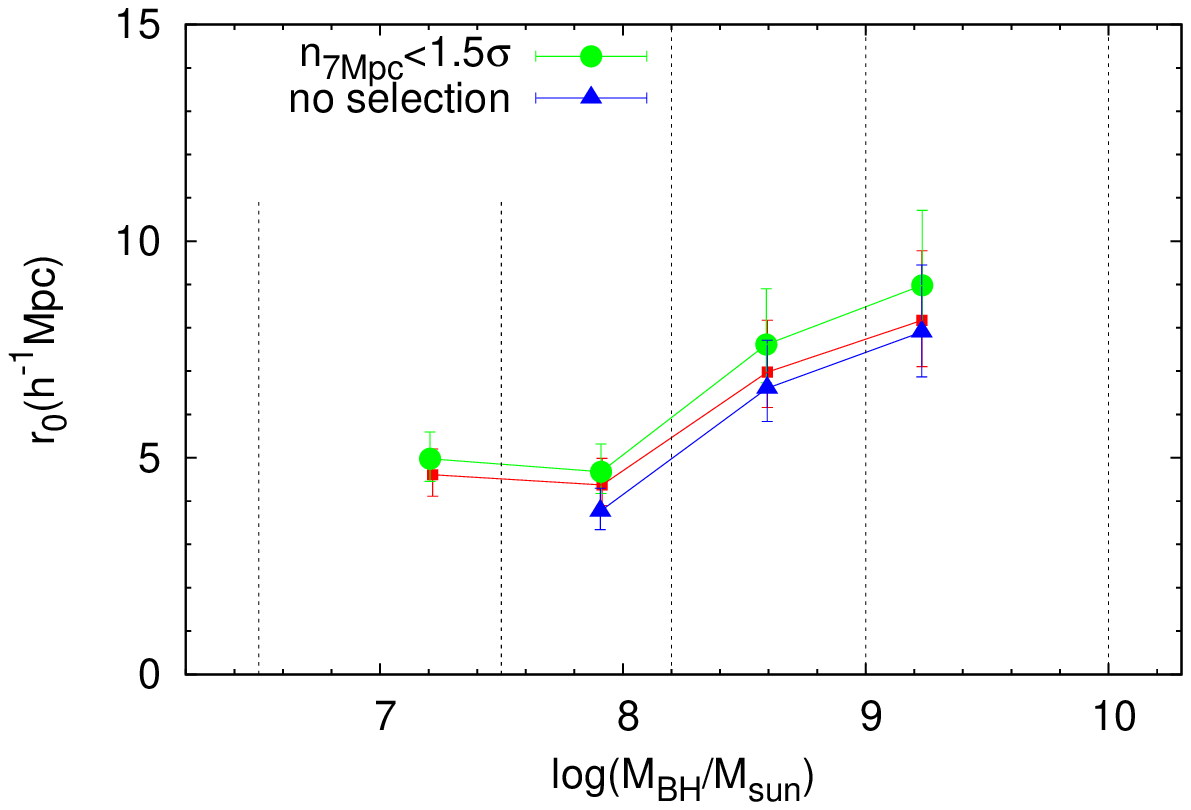}
\caption{ The scale length $r_0$ of cross-correlation for different selection criteria for AGN samples. 
{\it Blue triangles}: Results of the analysis without any sample rejection. 
{\it Red squares}: AGN samples with extremely inhomogeneous galaxy distribution around them are rejected by the criteria described in \citet{Shirasaki11} in order to reduce the effect of foreground clusters which are accidentally located near the sample AGNs on the sky. 
The same with Figure~\ref{m-r0}. 
{\it Green circles}: Samples with extremely high or low ($>1.5 \sigma$) projected number density of galaxies $n_{7 \rm Mpc}$ is rejected. 
See also Section~\ref{dataS}. 
 Slightly larger $r_0$ is estimated for the analysis with the additional selection. 
On the other hand, the all analysis gives almost same relative mass dependence. 
\label{fig:r0_criterion} 
}
\end{figure}

We rejected sample AGNs with anomalous galaxy distribution around the AGNs, as described in the Section~\ref{dataS} to remove the effect from foreground objects. 
To see whether we are removing real clustering by this rejection, we present the results of the analysis done both with and without the selection, in Figure~\ref{fig:r0_criterion}. 
The blue triangles show the results without any sample rejection. 
We cannot find significant clustering signal at $\log(\Mbh/\msun)<7.5$ for this analysis.  
The red squares are the fiducial results with the selection criteria described in \citet{Shirasaki11}. 
For green circles, we also reject AGN samples with extremely high or low projected number density ($n_{7 \rm Mpc}$) of galaxies around them. 

The analysis without selection gives slightly smaller $r_0$. 
This indicate that the clustering feature is weakened by foreground contamination for the analysis without sample rejection. 
The analysis with the $n_{7 \rm Mpc}$ criterion gives $r_0=6.3^{+0.9}_{-0.6}\hMpc$ for the whole mass range. 
This is $\sim 0.5 \hMpc$ larger than the result without the $n_{7 \rm Mpc}$ criterion. 
On the other hand, we see almost same relative mass dependence for the all three cases.

\subsection{AGN Redshift}
\begin{figure}
\includegraphics[width=\linewidth]{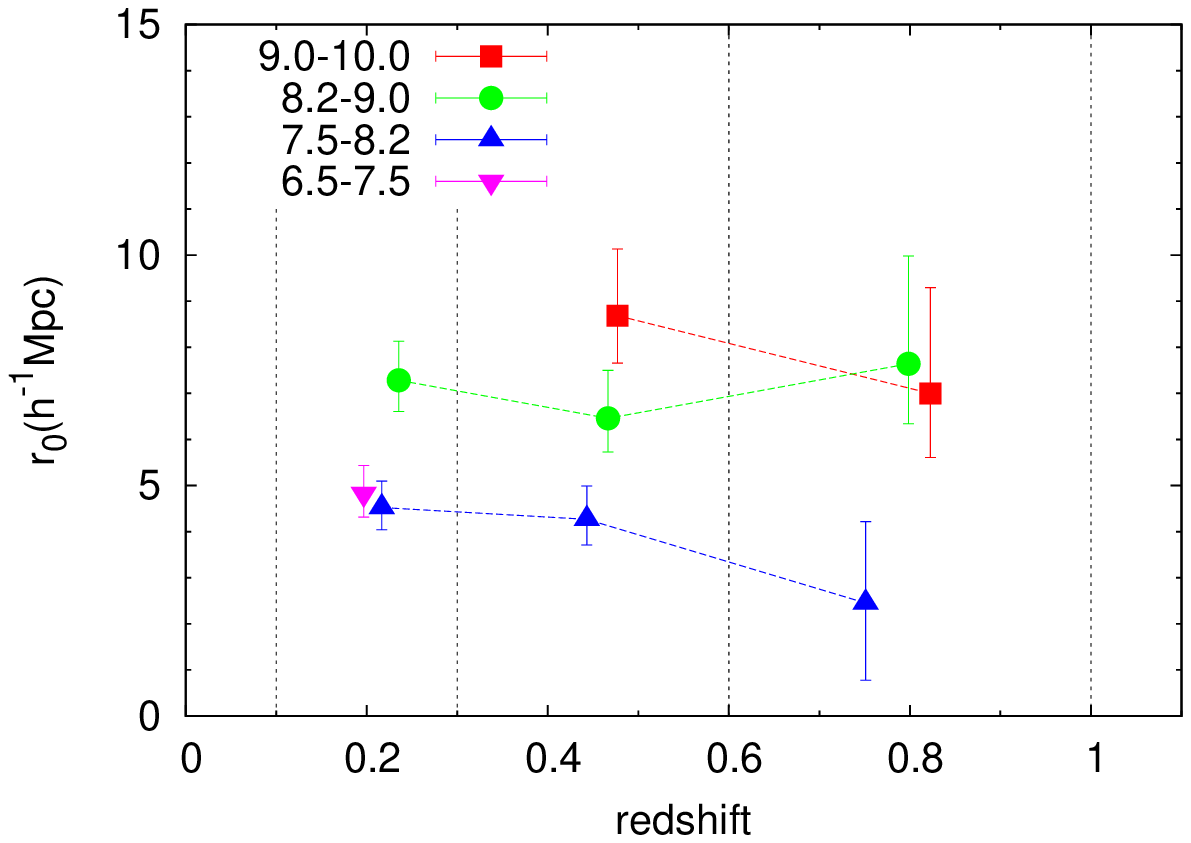}
\includegraphics[width=\linewidth]{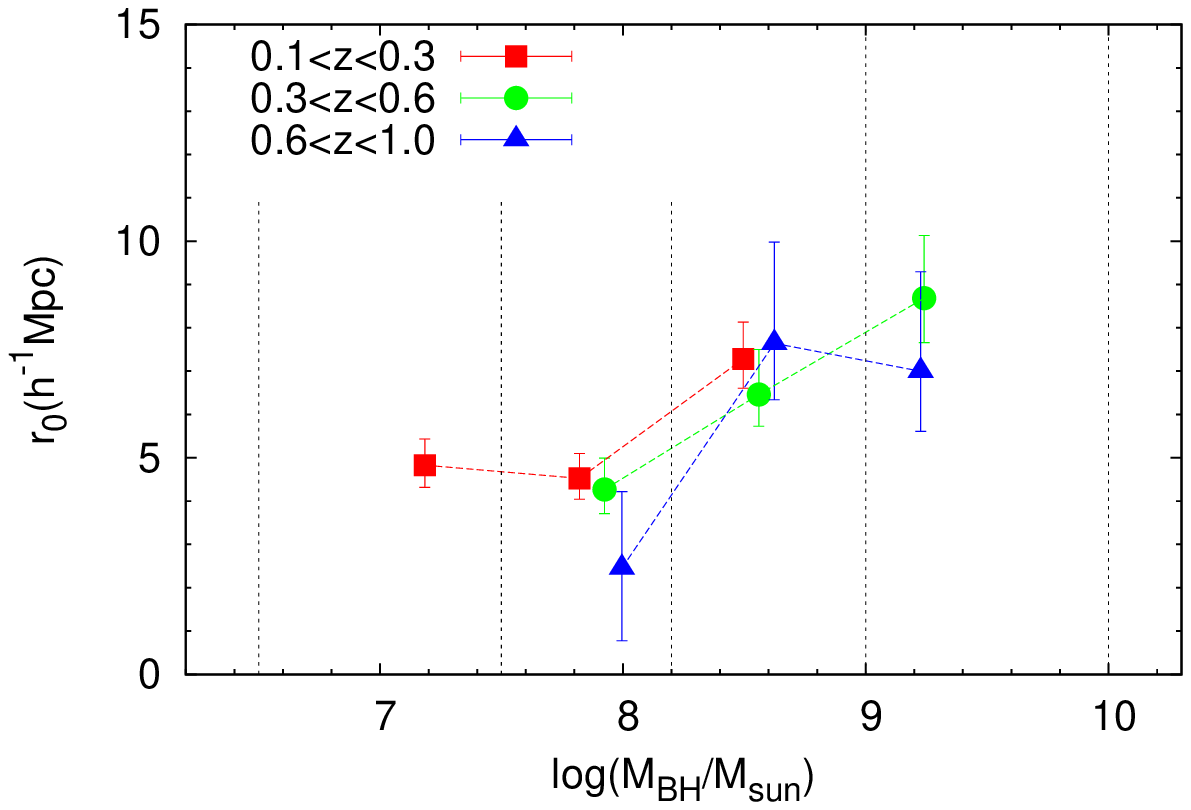}
\caption{
{\it Top panel}: The correlation length ($r_0$) against redshift for the samples of four BH mass ranges: 
$\log(\Mbh/\msun)=9.0$--$10.0$ (red squares), $8.2$--$9.0$ (green circles),  $7.5$--$8.2$ (blue triangles), and $6.5$--$7.5$ (magenta inverted triangles). 
Results of mass and redshift range with $n_{\rm AGN}<200$ are not plotted because of very large uncertainty. 
Vertical dotted lines show boundaries of redshift ranges.  
{\it Bottom panel}: $r_0$ against BH mass for the samples of three redshift ranges: 
$z=0.1$--$0.3$ (red squares), $z=0.3$--$0.6$ (green circles), and $z=0.6$--$1.0$ (blue triangles). 
}
\label{redshift}
\end{figure}

As seen in Figure~\ref{sample}, distribution of the BH mass in our sample depends on the redshift. 
This is because SMBHs with larger mass tend to be luminous and can be observed even at higher redshift. 
To see the dependence on the redshift, we divided the samples into three groups with redshift ranges of $z=0.1$--$0.3$, $z=0.3$--$0.6$, and $z=0.6$--$1.0$. 
Figure~\ref{redshift} shows the estimated $r_0$ for the three redshift ranges and four mass ranges. 
As seen in the top panel of this figure, $r_0$ is not dependent on the redshift. 
Therefore, the increasing trend of $r_0$ seen in Figure~\ref{m-r0} would not be due to the redshift bias. 
We could also see the mass dependence for the sub-samples of low redshift ($z=0.1$--$0.3$, red squares) and intermediate redshift ($z=0.3$--$0.6$, green circles). 
For higher redshift sample ($z=0.6$--$1.0$), the estimated error is too large to see the dependence on BH mass. 
We summarize the estimated $r_0$, $\langle \rho_0 \rangle$ and $\langle n_{\rm bg} \rangle$ for the each mass range and redshift range in Table~\ref{Tresult}. 

\begin{figure}
\plotone{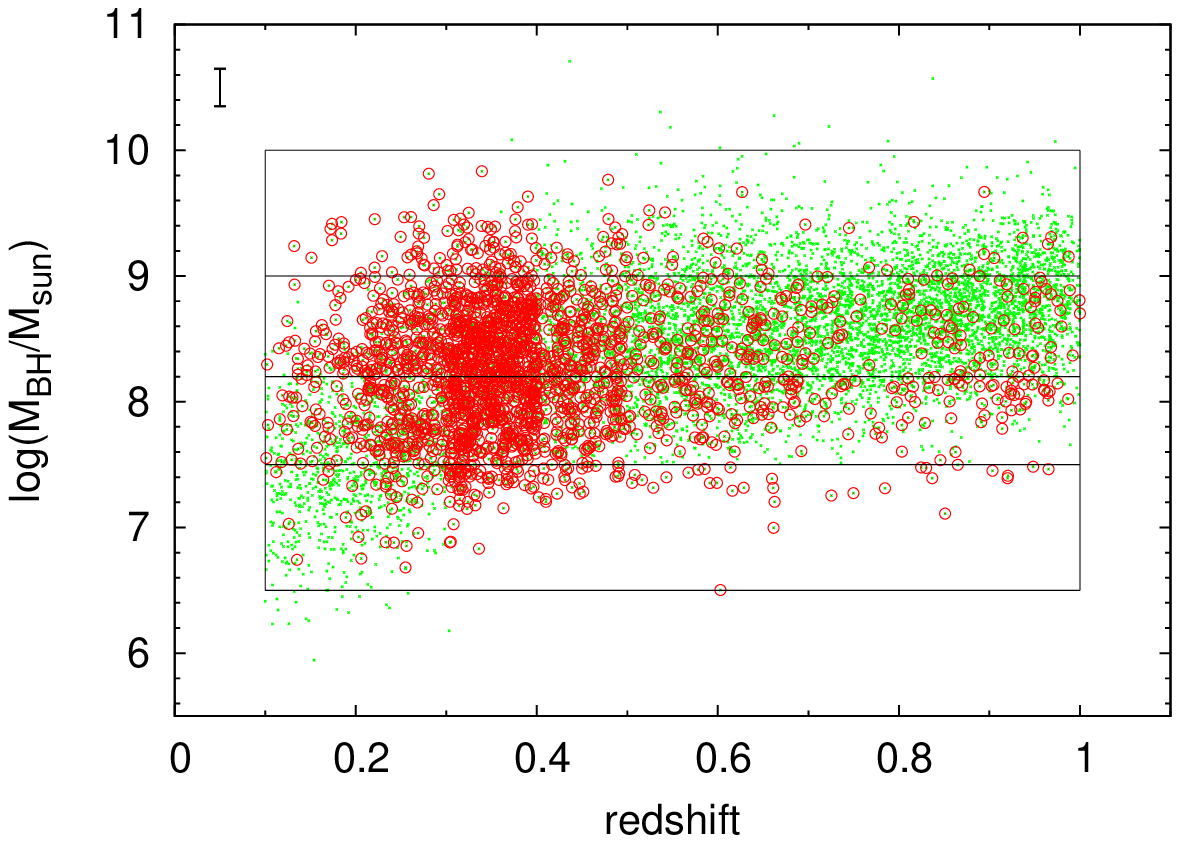}
\caption{Resampling of the AGN sample in order to cancel the redshift dependence of the BH mass. 
The green crosses are the original sample (shown in Figure~\ref{sample}) and the red circles denote one set of the resampled AGNs. 
Selection probability is determined to give the same redshift distribution for the four mass ranges. 
We construct ten sets of the resampled AGNs by the Monte-Carlo method. 
}
\label{subsample}
\end{figure}

\begin{figure}
\plotone{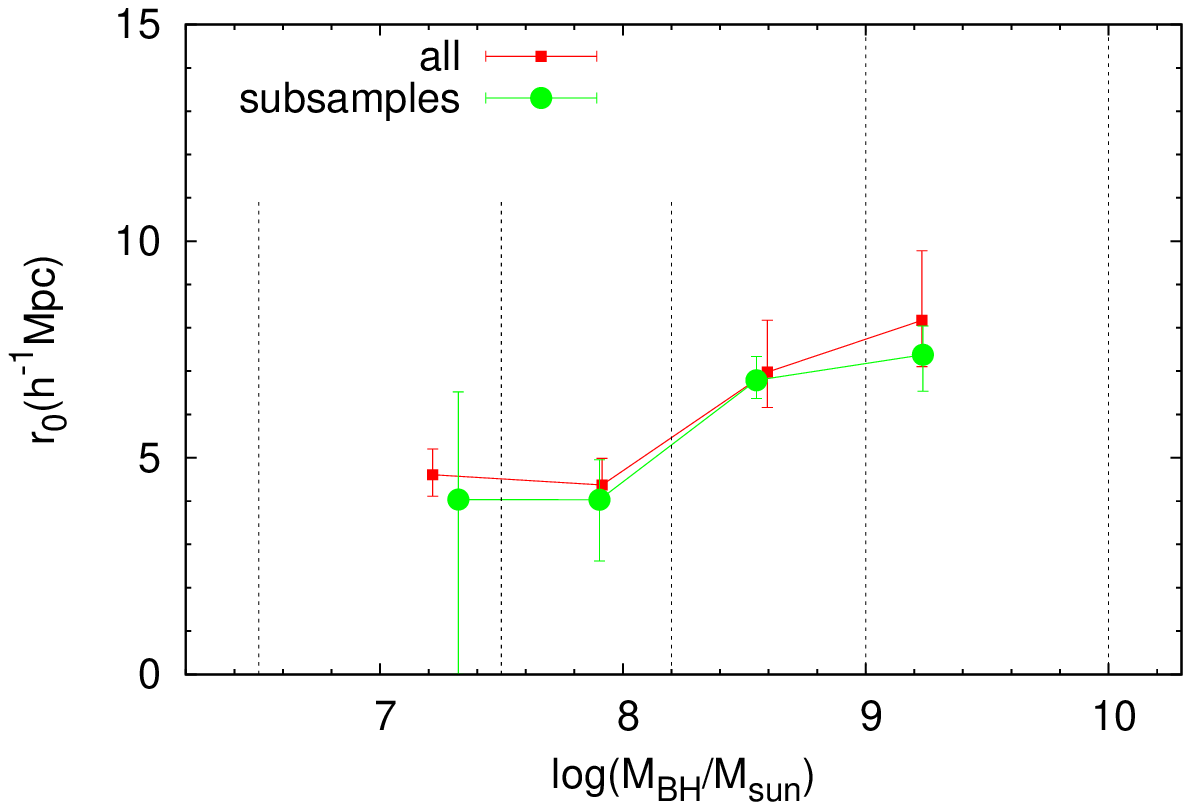}
\caption{The scale length $r_0$ of cross-correlation against BH mass for the resampled AGNs (green circles). 
The sub-samples is selected to cancel the redshift dependence of the BH mass. 
Symbols show median, and error bars show maximum and minimum of the ten sample sets constructed by the Monte-Carlo resampling. 
We see the similar trend as the analysis for the all sample AGNs (red squares). 
}
\label{m-r_subsample}
\end{figure}

To remove the possible redshift bias, we also present the mass dependence for sub-samples with the normalized redshift distributions. 
We constructed sub-samples as follows: 
For redshift bins with $\Delta z=0.1$, the selection probability is determined to give the same redshift distribution for the four mass ranges. 
We randomly selected AGN samples following the selection probability and constructed a sets of sub-samples. 
Figure~\ref{subsample} shows one example of the sub-sample. 
We constructed ten sets of sub-samples and measured $r_0$ for them. 
For these sub-samples, there is no correlation between redshift and BH mass. 

Figure~\ref{m-r_subsample} shows the mass dependence of $r_0$ for the resampled AGNs with normalized redshift distribution. 
We plot medians of the ten sets of sub-samples as green circles. Error bars show maximum and minimum values for the ten sets.  
The resampled AGNs also show similar mass dependence as Figure~\ref{m-r0}.  
Large mass SMBHs show strong clustering amplitude at $\Mbh>10^8\msun$. 
The clustering amplitude for AGNs with $\log(\Mbh/\msun)=6.5$--$7.5$ and $7.5$--$8.2$ is the almost same although the error bars are very large.  
The relative mass dependence for the sub-samples is free from redshift bias since the sub-samples for the four mass ranges have the same redshift distribution. 
We note that there is neither correlation between BH mass and luminosity for these sub-samples. 
Therefore, the BH mass dependence is thought to be neither due to redshift bias nor to luminosity bias. 

\subsection{AGN Luminosity}
\begin{figure}
\includegraphics[width=\linewidth]{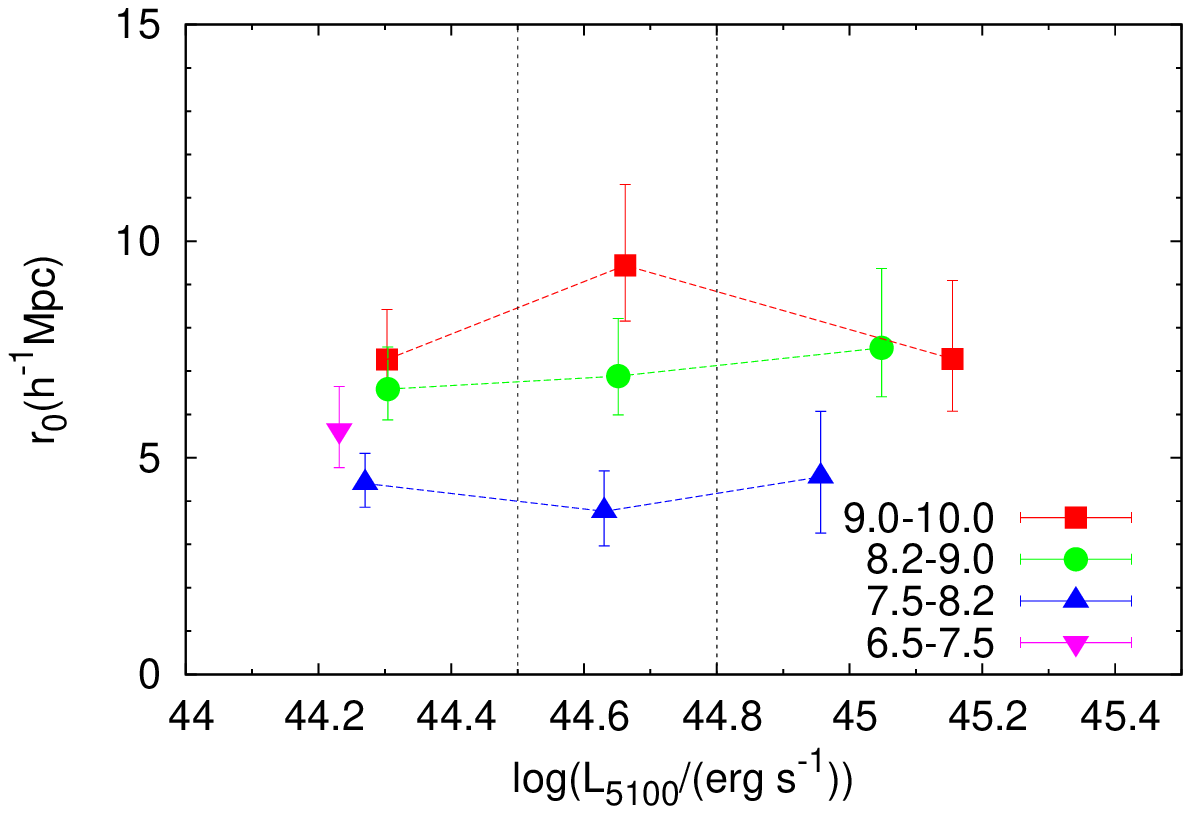}
\includegraphics[width=\linewidth]{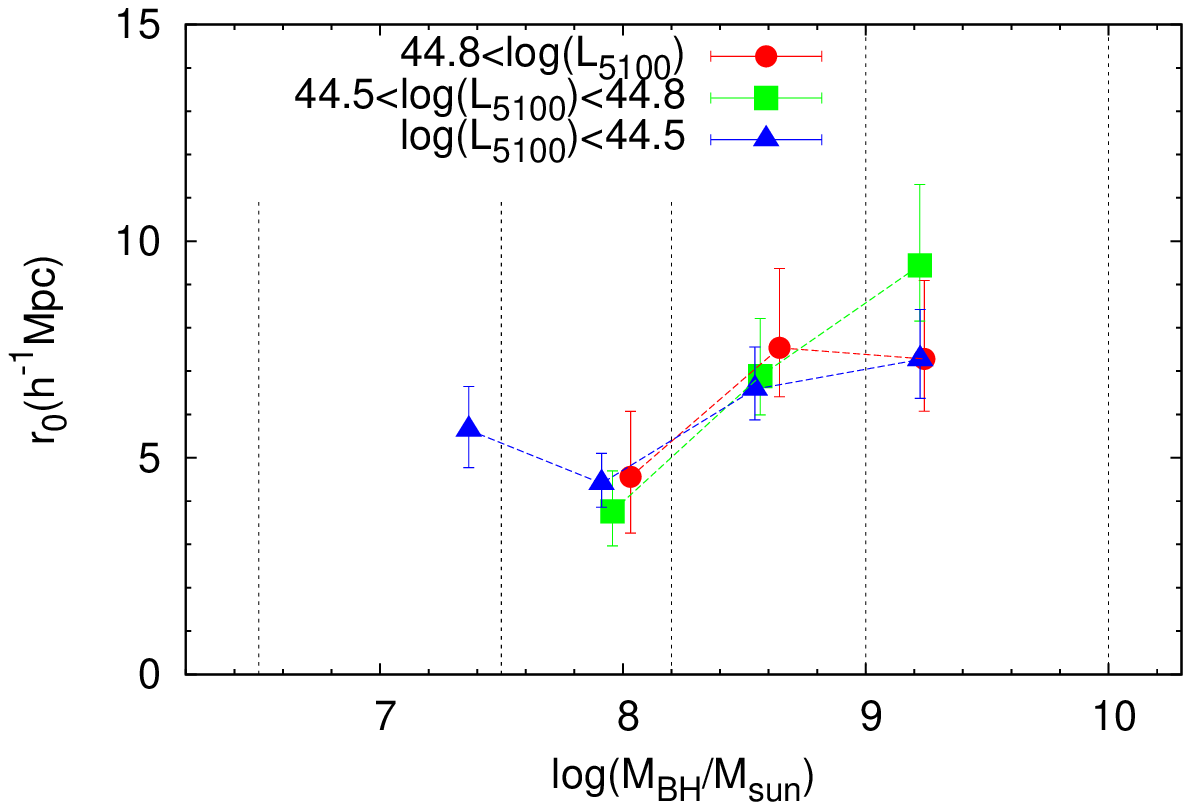}
\caption{
{\it Top panel}: The scale length ($r_0$) of cross-correlation against luminosity for the samples of four BH mass ranges: 
$\log(\Mbh/\msun)=9.0$--$10.0$ (red squares), $8.2$--$9.0$ (green circles),  $7.5$--$8.2$ (blue triangles), and $6.5$--$7.5$ (magenta inverted triangles). 
{\it Bottom panel}: $r_0$ against BH mass for the samples of three luminosity ranges: 
$L_{5100}\geq 10^{44.8}$erg/s (red circles), $10^{44.5}$erg/s $\leq L_{5100}<10^{44.8}$erg/s (green squares), and $L_{5100}<10^{44.5}$erg/s (blue triangles). 
}
\label{luminosity}
\end{figure}

We divided our sample into three luminosity range of $L_{5100}<10^{44.5}$erg/s,  $10^{44.5}$erg/s $\leq L_{5100}<10^{44.8}$erg/s, and  $10^{44.8}$erg/s $\leq L_{5100}$ to see luminosity dependence, where $L_{5100}$ is the monochromatic continuum luminosity at rest-frame 5100\AA. 
The top panel of Figure~\ref{luminosity} shows the dependence of $r_0$ on luminosity. 
We cannot find significant luminosity dependence for all the four mass ranges. 
On the other hand, we could see the mass dependence at $\Mbh>10^8\msun$ for all three luminosity ranges, as seen in the bottom panel.  
Therefore, we can conclude that the BH mass dependence seen in Figure~\ref{m-r0} is not due to the dependence on luminosity. 

\citet{Shen09} argued that the amplitude of AGN-AGN auto-correlation depends weakly on optical luminosity. 
\citet{Donoso10} found that the clustering amplitude varies with radio-luminosity on scales less than $\sim1$ Mpc but is almost independent on luminosity for the larger scale. 

These results indicate that the clustering amplitude on large scale depends on mass of SMBH but weakly depends on luminosity.

\subsection{Completeness and Luminosity of the Galaxy Sample}\label{completeS}

\begin{figure}
\plotone{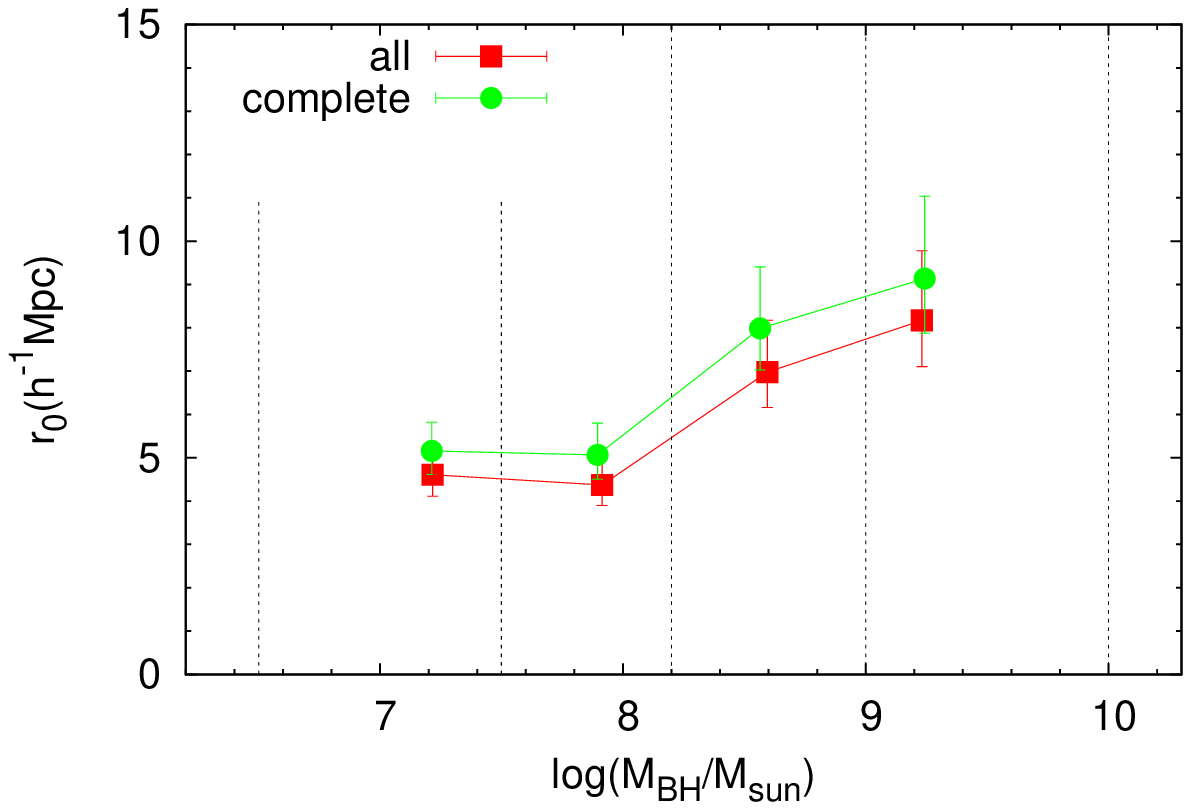}
\caption{The cross correlation length for complete galaxy sample as a function of virial mass 
of SMBH. 
The complete sample is galaxy samples which is blighter than threshold magnitude $\mth$ for 
each region around AGN. 
(See also Section \ref{analyS})
\label{fig:r0_complete} 
}
\end{figure}

\begin{figure}
\plotone{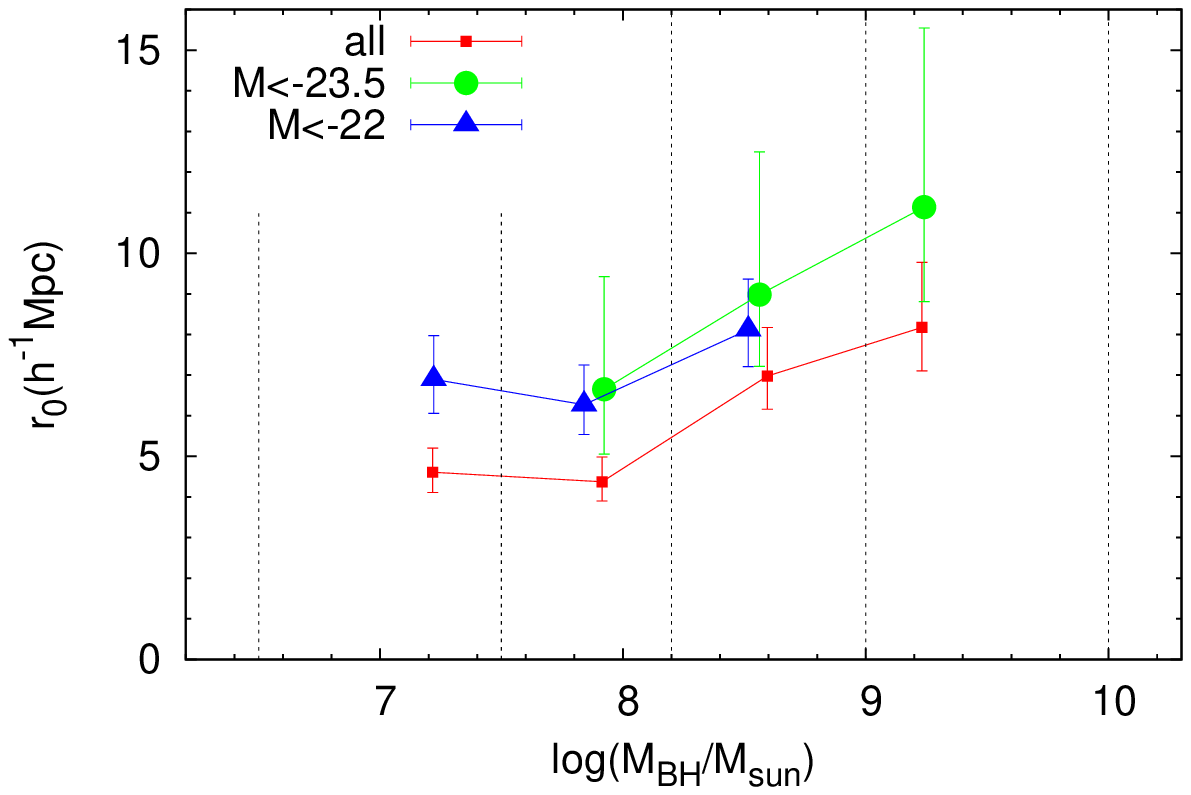}
\caption{The cross correlation length for luminosity limited galaxy samples as a function of virial
mass of SMBH. 
The luminosity limited samples are defined as 
$M \equiv m - {DM(z_{\rm AGN})} < -22.0$ (blue triangle) and $M < -23.5$ (green circle), respectively. 
Where $DM(z_{\rm AGN})$ represents distance modulus for the AGN redshift $z_{\rm AGN}$
\label{fig:r0_fluxlimited} 
}
\end{figure}

As described in Section~\ref{rho0S}, we correct incompleteness of the faint end of 
glalaxy sample by estimating the detection efficiency $DE(m)$ based on the magnitude 
distribution of UKIDSS sample for each area around an AGN. 
There may be a criticism that the correction of the incompleteness of the galaxy sample
is somehow biased to the luminosity of the galaxy sample. 
Figure~\ref{fig:r0_complete} shows the cross correlation length calculated for a complete 
galaxy sample which consist of blight galaxies with $m < m_{\rm th}$, where $m_{\rm th}$ is a threshold magnitude, below which $DE(m)=1$ (see Eq.~\ref{eq:det_eff}). 
For this analysis, $\rho_0$ is also recomputed by integrating the luminosity function upto $\mth$. 
This result also shows that the cross correlation increases above $M_{\rm BH} \sim 10^{8} \msun$. 
Therefore, the increasing trend is not due to the ambiguity that comes from using a incomplete galaxy sample.

It is known that brighter galaxies tend to cluster more strongly than dimmer galaxies. 
Thus it might be possible to explain the larger cross correlation length for more
massive SMBH by the bias due to the galaxy brightness.
Figure~\ref{fig:r0_fluxlimited} shows the cross correlation length calculated for 
luminosity limited samples. 
We selected the luminosity limited galaxy samples which are defined as 
$M \equiv m - {DM(z_{\rm AGN})} < -22$, and $M < -23.5$ for each AGN, where $DM(z_{\rm AGN})$ 
represents distance modulus for the AGN redshift. 
$M$ is not absolute magnitude for foreground and background galaxies but 
they should make no contribution to the clustering signal and not affect $r_0$. 
For these analysis, we selected AGN samples with $M_{\rm th} \equiv \mth - DM > -22.0$ and 
$M_{\rm th} > -23.5$, respectively (see also Figure \ref{fig:mlimit-abs-z}). 
Therefore, these absolute magnitude limited galaxy samples are also ``complete''($m < m_{\rm th}$). 
Although the error bar is relatively large, we can see the similar trend of the cross 
correlation length against virial mass of SMBH as Figure~\ref{m-r0}.
Significance of the difference of $r_0$ by considering statistical error is $1.9\sigma$ for mass ranges between $\log(\Mbh/\msun)=$7.5--8.2 and 8.2-9.0, and $1.7\sigma$ for between 8.2-9.0 and 9.0-10.0, for the analysis with  $M < -23.5$ sample, and $3.8\sigma$ for between $\log(\Mbh/\msun)=$7.5--8.2 and 8.2-9.0 for the $M < -22.0$ sample. 
Therefore, we can conclude that the increase of the cross correlation length seen in the
Figure~\ref{m-r0} is not only due to the bias related with the galaxy brightness.
The estimated scale length is larger than Figure~\ref{m-r0} since blighter galaxies are 
more strongly clustered. 

\section{Discussion}

In the previous studies, 
\citet{Shen09} have shown that most massive SMBHs are more strongly clustered than the remainders from auto-correlation analysis of quasars. 
It has also been shown that radio selected AGNs are strongly clustered than the cases for the optically selected AGNs \citep{Hickox09, Donoso10}, and characteristic BH mass of radio AGNs is higher than optically selected ones. 
Our results are consistent with these previous studies.  
These may indicate that the environment of galaxies has played an important role for the growth of high mass SMBHs. 
The clustering amplitude is relevant to the mass of the host dark-matter halo and the frequency of major merger. 
If mass growth of SMBH is mainly driven by the major mergers of galaxies, massive SMBHs are expected to be in massive halos. 

In contrast, we did not found significant luminosity dependence.  
Luminosity is thought to represent gas accretion activity at this time. 
Activity of SMBH is thought to be a transient event and not strongly correlate with large scale structure. 
On the other hand, black hole mass is thought to represent cumulative accretion history and merger history of BHs, and can be related with large scale environment. 

For less massive BHs with $\Mbh\lesssim 10^{8}\msun$, the significant correlation between $r_0$ and $\Mbh$ is not seen in our study. 
BH mass has been thought to be correlated with the mass of dark matter halo \citep[e.g.][]{Ferrarese02}. 
Recently, however, \citet{Kormendy11} found that the mass of SMBH does not directly correlate with dark matter halo mass, at least for low-mass SMBHs in disk galaxies, 
based on the observations of nearby SMBHs for which mass of the host dark halos are derived by the stellar kinematics. 

One possible scenario to explain the absence of the positive mass dependence for $\Mbh\lesssim 10^{8}\msun$ would be that the less massive BHs could be formed in the isolated galaxies by secular processes.  
If they have grown by secular processes, mass of a seed BH should be much larger than a typical stellar mass BH \citep[for a review, see, e.g.,][]{Volonteri10}.
Some authors \citep{Lodato06, Begelman06} argue that SMBHs with $10^4$--$10^6\msun$ are formed through the direct collapse of pre-galactic gas at $z>$10. 
Such a heavy seed BH can grow to $\sim10^7$--$10^8\msun$ by a few Gyrs without BH merger under the assumption that the mass accretion rate is around $\sim0.1$ times the Eddington rate. 
Another scenario would be that they are in a growing phase by the major mergers of galaxies. 

For the clustering amplitude of less massive SMBHs, the contribution of the AGNs hosted in satellite galaxies in the massive dark-matter halos can also be important, while most massive SMBHs are in the central regions of dark-matter halos.  
\citet{Padmanabhan09} argue that the satellite fraction of quasars is more than $25\%$ in their sample, which is selected from the SDSS catalog. 
The percentage should be larger for the less massive BHs. 
Further investigation of less massive BHs is crucial to understand formation and evolution mechanisms of a SMBH.

\begin{table*}
\caption{Statistics of fitting parameters for each virial mass and redshift group}
\label{Tresult}
\begin{center}
\begin{tabular}{|ll|ccc|rrr|}
\hline
virial mass & redshift & $n_{\rm AGN}$\tablenotemark{a}
 & $\langle \log(\Mbh/\msun) \rangle $\tablenotemark{b}
 & $\langle z \rangle $\tablenotemark{c}
 & $r_0$\tablenotemark{d}
 & $ n_{\rm bg} $\tablenotemark{e}
 & $\langle \rho_{0} \rangle $\tablenotemark{f} \\
 $\log(\Mbh/\msun)$ & & & &  & $\hMpc$ & $\Mpc^{-2}$ & $10^{-3}\Mpc^{-3} $ \\
\hline
 all	 & 0.1--1.0 & 9394 & 8.42 & 0.59 & $5.8^{+0.8}_{-0.6}$ & 10.473$\pm$0.005 & 1.9$\pm$0.4\\
\hline
$9.0$--$10.0$ 	& 0.1--1.0 &  1331 & 9.23 & 0.72 & $8.2^{+1.6}_{-1.1}$ & 4.542$\pm $0.008 & $0.93 \pm 0.25$ \\
 & 0.1--0.3 &   33 & 9.27 & 0.24 & --\tablenotemark{g} & -- & --    \\
 & 0.3--0.6 &  347 & 9.24 & 0.48 & $8.7^{+1.4}_{-1.0}$ & 7.29$\pm $0.02 & $2.0 \pm 0.5$    \\
 & 0.6--1.0 &  951 & 9.22 & 0.82 & $7.0^{+4.9}_{-2.1}$ & 2.70$\pm $0.01 & $0.38 \pm 0.15$    \\
 \hline
$8.2$--$9.0$ 	& 0.1--1.0 & 5119 & 8.60 & 0.66 & $7.0^{+1.2}_{-0.8}$ & 6.001$\pm $0.005 & $1.3 \pm 0.3$ \\
 & 0.1--0.3 &  320 & 8.50 & 0.24 & $7.3^{+0.9}_{-0.7}$ & 29.40$\pm $0.04 & $5.3 \pm 0.9$    \\
 & 0.3--0.6 & 1635 & 8.56 & 0.47 & $6.5^{+1.0}_{-0.7}$ & 7.49$\pm $0.01 & $2.1 \pm 0.5$    \\
 & 0.6--1.0 & 3164 & 8.62 & 0.80 & $7.6^{+2.3}_{-1.3}$ & 2.87$\pm $0.004 & $0.42 \pm 0.16$    \\
 \hline
$7.5$--$8.2$ 	& 0.1--1.0 & 2278 & 7.91 & 0.44 & $4.4^{+0.6}_{-0.5}$ & 15.69$\pm $0.01 & $2.8 \pm 0.6$ \\
 & 0.1--0.3 &  664 & 7.82 & 0.22 & $4.5^{+0.6}_{-0.5}$ & 38.26$\pm $0.03 & $5.7 \pm 1.0$    \\
 & 0.3--0.6 &  967 & 7.92 & 0.44 & $4.3^{+0.7}_{-0.6}$ & 8.53$\pm $0.01 & $2.3 \pm 0.5$    \\
 & 0.6--1.0 &  642 & 8.00 & 0.75 & $2.5^{+1.8}_{-1.7}$ & 3.22$\pm $0.01 & $0.53 \pm 0.19$    \\
 \hline
$6.5$--$7.5$ 	& 0.1--1.0 & 635 & 7.22 & 0.25 & $4.6^{+0.6}_{-0.5}$ & 38.65$\pm $0.03 & $5.4 \pm 0.9$ \\
 & 0.1--0.3 &  513 & 7.18 & 0.20 & $4.8^{+0.6}_{-0.5}$ & 45.69$\pm$0.04 & $6.1 \pm 1.0$    \\
 & 0.3--0.6 &  101 & 7.36 & 0.38 & -- & -- & -- \\ 
 & 0.6--1.0 &   21 & 7.30 & 0.77 & -- & -- & -- \\
 \hline
\end{tabular}
\tablenotetext{a}{number of sample AGNs}
\tablenotetext{b}{average of logarithm of BH mass}
\tablenotetext{c}{average redshift}
\tablenotetext{d}{correlation length, the error contains the systematic error due to the uncertainty of $\rho_0$ and the $1\sigma$ statistical error.}
\tablenotetext{e}{average of the projected number density of background galaxies}
\tablenotetext{f}{average of the averaged number density of galaxies at the AGN redshift}
\tablenotetext{g}{We do not derive parameters for sub-sample with $n_{\rm AGN}<200$}
\end{center}
\end{table*}

\section{Conclusions}
We have investigated the clustering of galaxies around 9,394 AGNs for $z=0.1$--$1$. 
We obtained the galaxy data of UKIDSS LAS by means of virtual observatory tools. 
Our results are free from the effect of cosmic variance owing to the large sample covering the large area of the sky. 
The estimated correlation length ranges between $4$--$10\hMpc$ depending on BH mass, and depends neither on redshift for $z=0.1$--$1$ nor on luminosity. 
The results may indicate that higher mass BHs reside in more clustered environment for $\Mbh>10^8\msun$. 

While our results show positive mass dependence for $\Mbh>10^8\msun$, our results for $\Mbh\lesssim 10^{8}\msun$ show no significant mass dependence. 
Although our sample of less massive BHs is small, this would give a critical mass scale for emergence of environment effect for BH growth. 

In this study the redshift range where the BH mass dependence of the cross correlation is measured with good accuracy is limited to below $z\sim0.6$. 
This is because the number density of UKIDSS LAS galaxies is too low at higher redshift, and also the systematic error due to the uncertainty of $M_*$ parameter of luminosity function is too large.
To understand the relation of the BH mass accretion history and it environment, it is crucial to observe its evolution upto at least redshift two where the number density of QSO is maximal and the mass accretion is expected to be the most prosperous.
To reduce the uncertainty, it is crucial to perform deeper observations so that the limiting magnitude reaches well beyond the characteristic luminosity of galaxies at AGN redshifts. 
At redshifts larger than 0.5, the dominant factor to the uncertainty of $\rho_{0}$ is
uncertainty of $M_{*}$ parameterization and the $\rho_{0}$ uncertainty becomes larger
than 20\%.
Two magnitude deeper observation will extend the redshift ranage where the uncertainty of 
$\rho_{0}$ is less than 20\% up to 1.0.
Future instruments such as Hyper Suprime-Cam (HSC) can measure the AGN environment more accurately
with good statistics.
When the survey is performed with 26 mag in $r$ band with HSC, we can estimate $\rho_{0}$
with accuracy less than 20\% up to redshift 2, and as a result can estimate $r_{0}$ with 
accurary less than 10\%.
Such a deep and wide survey would reveal the mechanism of AGN evolution at an important
epoch that its activity was the most prosperous.
\vspace{0.5cm}
\\

We appreciate to S. Eguchi and M. Enoki for their useful discussions. 
Results are based on data obtained from the Japanese Virtual Observatory, which is operated by the Astronomy Data Center, National Astronomical Observatory of Japan. 
This research has made use of the VizieR catalogue access tool, CDS, Strasbourg, France. 
This work is based on data obtained as part of the UKIRT Infrared Deep Survey. 
Funding for the SDSS and SDSS-II has been provided by the Alfred P. Sloan Foundation, the Participating Institutions, the National Science Foundation, the U.S. Department of Energy, the National Aeronautics and Space Administration, the Japanese Monbukagakusho, the Max Planck Society, and the Higher Education Funding Council for England. The SDSS Web Site is http://www.sdss.org/.


\end{document}